\documentclass[12pt]{article}
\usepackage{amsmath}
\usepackage{amsfonts}
\usepackage{amssymb}
\usepackage{graphicx}
\usepackage{hyperref}
\usepackage{multirow}
\usepackage{authblk}
\usepackage{color}
\usepackage[dvipsnames]{xcolor}

\usepackage{amsthm}
\usepackage{latexsym}
\usepackage{lscape}
\usepackage{natbib}

\newtheorem{theorem}{Theorem}[section]

\newtheorem{remark}[theorem]{Remark}

\newcommand{\ignore}[1]{}

\renewcommand{\epsilon}{\varepsilon}
{\begingroup

 \begin{enumerate}}
 {\end{enumerate}\endgroup}

\newcommand{\subfigimg}[3][,]{%
	\setbox1=\hbox{\includegraphics[#1]{#3}}
	\leavevmode\rlap{\usebox1}
	\rlap{\hspace*{-5pt}\raisebox{\dimexpr\ht1-2\baselineskip}{#2}}
	\phantom{\usebox1}
}
\title{Mixed-mode bursting oscillations in a three-timescale biophysical neuronal oscillator model}
\author[1]{Ngoc Anh Phan}
\author[1]{Yangyang Wang}

\affil[1]{\footnotesize Department of Mathematics, Brandeis University, Waltham, Massachusetts 02453, USA}

\begin{document}
\maketitle
\begin{abstract}
    Mixed-mode oscillations (MMOs), characterized by the alternation of small-amplitude oscillations (SAOs) and large-amplitude oscillations (LAOs), and bursting oscillations are common forms of complex oscillatory dynamics observed in systems with multiple timescales and have been widely studied across scientific disciplines. Mixed-mode bursting oscillations (MMBOs) combine features of MMOs and bursting, with LAOs organized into burst events. Most existing studies treat MMBOs as two-timescale phenomena, identifying distinct geometric mechanisms depending on how the timescales are grouped. In this work, we use a three-timescale biophysical cortical neuronal oscillator model to demonstrate that a three-timescale implementation of geometric singular perturbation theory (GSPT) provides stronger predictive insight into MMBO dynamics. This perspective unifies mechanisms previously identified from fast-slow analysis, while revealing the canard-delayed-Hopf (CDH) singularity as an organizing center for MMBOs near the singular limit. We perform a detailed bifurcation analysis of the full eight-dimensional model to determine how MMBOs are organized along families of isolas. We then combine GSPT with full-system bifurcation analysis to show how tuning the relative timescales induces transitions among MMOs, MMBOs, and bursting dynamics. Our results highlight the importance of studying MMBOs from a three-timescale perspective and demonstrate that combining GSPT with full-system bifurcation analysis can reveal organizing structures and mechanisms that are not apparent from either approach alone.
\end{abstract}

\section{Introduction}\label{sec:intro-theta}


Bursting oscillations are commonly observed in multiple-timescale systems and have been widely studied across scientific disciplines, including physics \cite{Kingni2015, Ma2022, Leutcho2020a, Leutcho2020b}, chemistry \cite{Rinzel1982, Beims2018, Organ2003}, and neuroscience \cite{Rinzel2006, Desroches2022, Rinzel1987, Izhikevich2000, Bertram1995, Golubitsky2001, Ermentrout2010, RE1998, Hindmarsh1984, Bertram2008, Kepecs2000, WR2016, WR2017, WR2020, Avitabile2022}. 
These dynamics alternate between silent and active phases, with the active phase consisting of multiple spikes. They are typically analyzed through a fast-slow decomposition \cite{Fenichel1979,Rinzel1987}, in which the slow variables are treated as parameters for the fast subsystem, and the bifurcation structure organizes the onset and termination of bursts \cite{Rinzel1987, Izhikevich2000, Bertram1995, Golubitsky2001}.


Another common type of complex oscillatory dynamics in systems with multiple timescales is mixed-mode oscillations (MMOs), characterized by the alternation of small-amplitude oscillations (SAOs) and large-amplitude oscillations (LAOs) within each periodic cycle \cite{Desroches2012}. Like bursting dynamics, MMOs have been recognized in many branches of sciences including physics, chemistry and particularly life sciences such as \cite{Hudson1979,Awal2023,Krupa2008a,Yu2008,Krupa2012,Teka2012, Vo2010, Vo2014,Curtu2010,CR2011,Harvey2011,Kugler2018,Kimrey2020,2ndKimrey2020,Pavlidis2022,Bat2021}. 
Theoretical analysis of MMOs has been well developed through geometric singular perturbation theory (GSPT) \cite{Fenichel1979}; see \cite{Desroches2012} for review. 
In two-timescale systems, two widely studied MMO mechanisms are canard dynamics associated with the twisting of slow manifolds due to folded singularities \cite{SW2001,Wechselberger2005} and slow passage through a delayed Andronov--Hopf bifurcation (DHB) of the fast subsystem \cite{Baer1989, Neishtadt1987, Neishtadt1988, Hayes2016,engler2026delays}. In three-timescale systems, these mechanisms need not remain separated. Instead, they can coexist and interact near a canard-delayed-Hopf (CDH) singularity \cite{Teka2012, Vo2013, Maess2014,Letson2017,Phan2023, he2026complex}. In \cite{Phan2023}, we showed that CDH singularities can organize robust MMO dynamics and that changing timescale parameters can shift the relative influence of the DHB and folded-node canard mechanisms.

Mixed-mode bursting oscillations (MMBOs), also called folded-node bursting oscillations, combine features of bursting and MMOs \cite{Desroches2013}. They are analogous to MMOs, but with the LAOs organized into burst events. In continuous dynamical systems, this typically requires at least two fast variables to support continuous spiking within bursts in the fast subsystem and at least two slow variables to support canard-mediated SAOs in the slow subsystem. First described in \cite{Desroches2013}, MMBOs were shown to be organized by folded-node canards, with both SAOs and spike-adding of the underlying burster mediated by canards \cite{DK2018, Nowacki2012}. Subsequent work has analyzed MMBOs in simplified two-timescale models, including Hindmarsh--Rose-type polynomial vector fields \cite{Desroches2013,Desroches2022}, rate models \cite{koksal2020canard}, and integrate-and-fire-type models \cite{Rubin2017,Yu2021}. In particular, \cite{koksal2020canard} showed that MMBOs are organized along isolated bifurcation branches in parameter space, known as isolas \cite{fernandez1997isolas}, along which spike-adding transitions occur. Such isola structures are also commonly observed in fast-slow dynamical systems exhibiting bursting, chaotic bursting \cite{tsaneva2010full, Farjami2020, barrio2020spike, barrio2024dynamics} or MMO-type complex oscillations \cite{vo2012bifurcations,vo2026symmetric}.

 

Despite this progress, MMBOs remain much less understood than bursting or MMOs, particularly in systems with more than two timescales, yet many systems exhibiting MMBO-type dynamics involve more than two distinct timescales \cite{Desroches2022, Pittman2021}.
Different reductions of the same three-timescale model to two-timescale problems can emphasize different mechanisms \cite{Letson2017,Phan2023}. For example, \cite{Bertram2008, Desroches2022} examined MMBOs in an episodic burster model with three distinct timescales. 
Freezing the slowest of the two slow variables reveals a fast-subsystem DHB mechanism \cite{Bertram2008}, whereas treating the two slow variables on the same timescale highlights folded-node canard dynamics as a more robust mechanism explaining the observed MMBO pattern \cite{Desroches2022}. 
While each reduction provides valuable insight, it does not fully address how these mechanisms coexist, interact, or exchange dominance when all three timescales are retained. This motivates a fully three-timescale approach, in which the relative separation between the fast, slow, and superslow variables can be varied explicitly to determine how DHB, canard, and other mechanisms jointly organize MMBO dynamics.


The focus of this paper is the investigation of these mixed-mode bursting patterns in a fully three-timescale setting by considering a biophysical conductance-based model of cortical theta neuronal oscillators introduced in \cite{Pittman2021} in the context of speech perception. 
After nondimensionalization, the model takes the following general form
\begin{equation}\label{eq:general-three-timescale}
\begin{aligned}
\varepsilon \dot{x} &= f(x,y,z), \\
\dot{y} &= g(x,y,z), \\
\dot{z} &= \delta h(x,y,z),
\end{aligned}
\end{equation}
where $0<\varepsilon,\delta\ll 1$ are independent perturbation parameters, $x$ denotes fast variables, $y$ denotes slow variables, and $z$ denotes superslow variables. 
This structure allows us to apply extended GSPT \cite{Fenichel1979, Vo2013, Nan2015}, explicitly accounting for all three timescales. 
We are interested in understanding how varying the separation of timescales controls the behavior of the system, with a focus on MMBO-type dynamics and their transitions to MMOs and bursting dynamics. 
We begin by performing the GSPT analysis of the three-timescale model \eqref{eq:general-three-timescale} and show that the singular orbits provide reliable predictions for the full-system periodic bursting solutions for $0<\varepsilon,\delta\ll 1$ small enough (Fig.~\ref{fig:singular-orbit}). We then go beyond the singular limits to examine bifurcations of the full system with respect to the speed of the fast variable and the speed of the superslow variable over broader ranges of $\varepsilon$ and $\delta$.

The main contributions of the present study lie in several aspects. First, we provide, to the best of our knowledge, the first detailed analysis of MMBOs in a fully three-timescale dynamical system with two independent singular perturbation parameters $\varepsilon$ and $\delta$. 
In this setting, we compute the isolas of MMBOs in an eight-dimensional conductance-based neuronal model and show that, unlike the MMBO isolas in a two-slow/two-fast model \cite{koksal2020canard}, these isolas can contain multiple stable segments associated with different spike numbers and are strongly shaped by the timescale parameters.
Second, we demonstrate that SAOs in the same model can arise through multiple mechanisms, including DHB, folded-node canard effects, CDH, saddle-focus dynamics of the full system, and interactions among them. Finally, by varying $\varepsilon$ and $\delta$ independently, we show how timescale structure selects among these mechanisms and thereby controls transitions between MMOs, MMBOs, and regular bursting.

The remainder of this paper is outlined as follows. We introduce the full theta oscillator model \eqref{eq:main-theta} in Section \ref{sec:model}. In Section \ref{sec:gspt-theta}, we present a geometric singular perturbation analysis of \eqref{eq:main-theta}, treating $\varepsilon$ and $\delta$ as independent perturbation parameters. 
This analysis clarifies how the DHB and folded node are related to the CDH in the double singular limit and allows us to construct singular orbits that predict the full-system dynamics for the $0<\varepsilon,\delta\ll 1$ regime. 
In Section \ref{sec:numerical-simu-theta}, we go beyond the singular limit analysis by using full-system bifurcation analysis to follow periodic solutions over broader ranges of $\varepsilon$ and $\delta$. This reveals that MMO and MMBO solutions are organized along isolas, whose stable segments and bifurcations mediate spike-adding and transitions between activity patterns. In Section \ref{sec:mmos-mmbos-transition}, we link these two viewpoints to explain how different transitions occur.  Finally, we conclude with a discussion in Section \ref{sec:discussion-theta}.

\section{The mathematical model}\label{sec:model}

We consider an eight-dimensional (8D) conductance-based cortical theta oscillator model \cite{Pittman2021}. This model contains three inward currents, namely fast sodium ($I_{\mathrm {Na}}$), persistent sodium ($I_{\mathrm {NaP}}$), and calcium ($I_{\mathrm {Ca}}$) currents, and four outward currents, namely delayed-rectified potassium ($I_{\mathrm {K_{DR}}}$), leak ($I_{\mathrm {leak}}$), potassium ($I_m$), and calcium-dependent potassium currents ($I_{\mathrm {K_{SS}}}$). The dynamics of voltage ($V$), activation variables, and intracellular calcium concentration ($\mathrm {\mathrm{Ca_i}}$)  are governed by the following equations:

\begin{equation}\label{eq:main-theta}
\begin{array}{rcl}
     C\frac{dV}{dt}&=& I_{\mathrm {app}} - I_{\mathrm {Na}}-I_{\mathrm {K_{DR}}}-I_{\mathrm {leak}}-I_m -I_{\mathrm {NaP}}-I_{\mathrm {Ca}}-I_{\mathrm {K_{SS}}},\\
     \frac{dm_{\mathrm {NaP}}}{dt}&=& (m_{\infty}(V)-m_{\mathrm {NaP}})/\tau_{m},\\
     \frac{ds}{dt}&=& (1-s)\alpha_s-s\beta_s,\\
     \frac{dm_{\mathrm {K_{DR}}}}{dt}&=& \tau_{\mathrm {fast}}((1-m_{\mathrm {K_{DR}}})\alpha_{m}-{m_{\mathrm {K_{DR}}}}\beta_{m}),\\
     \frac{dh}{dt}&=& \tau_{\mathrm {fast}}((1-h)\alpha_h-h\beta_h),\\
     \frac{d\mathrm {\mathrm{Ca_i}}}{dt}&=& -F_{\mathrm {Ca}} I_{\mathrm {Ca}}(V,s)-\mathrm {\mathrm{Ca_i}}/\tau_{\mathrm {Ca}},\\
      \frac{dn}{dt}&=& (n_{\infty}(V)-n)/\tau_n(V),\\
     \frac{dq}{dt}&=& (1-q)\alpha_q( {\mathrm {\mathrm{Ca_i}}})-q\beta_q,
\end{array}    
\end{equation}
where the currents are given by
\begin{equation}
    \begin{array}{rcl}
       I_{\mathrm {Na}}  &=& g_{\mathrm {Na}}m_{\mathrm {Na}}(V)^3h(V-E_{\mathrm {Na}})  \\
       m_{\mathrm {Na}}(V)&=&\alpha_{m_{\mathrm {Na}}}(V)/(\alpha_{m_{\mathrm {Na}}}(V)+\beta_{m_{\mathrm {Na}}}(V))\\
       I_{\mathrm {K_{DR}}}  &=& g_{\mathrm {K_{DR}}}m^4_{\mathrm {K_{DR}}}(V-E_K)\\
       I_{\mathrm {leak}}&=& g_{\mathrm {leak}} (V-E_{\mathrm {leak}})\\
       I_m &=& g_m n(V-E_K)\\
       I_{\mathrm {NaP}} &=& g_{\mathrm {NaP}} m_{\mathrm {NaP}}(V-E_{\mathrm {NaP}})\\
       I_{\mathrm {Ca}} &=& g_{\mathrm {Ca}}s^2(V-E_{\mathrm {Ca}})\\
       I_{\mathrm {K_{SS}}} &=& g_{\mathrm {K_{SS}}}q(V-E_K).\\
    \end{array} 
\end{equation}

We set the applied current $I_{\mathrm {app}}$ to be 8 $\mathrm{\mu A/cm^2}$, placing the system in the mixed-mode oscillation (MMO) regime. Other parameter values are the same as in \cite{Pittman2021}, with corresponding units and activation variable dynamics listed in Tables \ref{tab:para-theta} and \ref{tab:activation} in Appendix \ref{sec:para-theta}.  
In addition to MMOs, the model is capable of generating mixed-mode bursting oscillations (MMBOs) and regular bursting dynamics, in consistency with \emph{in vitro} recordings from layer 5 theta-resonant pyramidal cells \cite{carracedo2013neocortical}.
To facilitate our analysis, we first perform a dimensional analysis of the full system \eqref{eq:main-theta}. The details of the nondimensionalize procedure are given in Appendix \ref{sec:nondim-theta}, which transforms \eqref{eq:main-theta} to:
\begin{equation}\label{eq:slow-theta}
\begin{array}{rcl}
 \varepsilon_1\frac{dV}{dt_s} &=&  f_1(V,\mathbf{y}, n,q),\vspace{0.05in}\\
 \varepsilon_i\frac{dy_i}{dt_s} &=&  f_i(V,\mathbf{y}),\vspace{0.05in}\\
  \frac{dn}{dt_s}&=& g_1(V,n),\vspace{0.05in}\\
    \frac{dq}{dt_s}&=& \delta  g_2(\mathrm{\mathrm{Ca_i}},q),
\end{array}
\end{equation}
where $\mathbf{y} = [\,m_{\mathrm{NaP}},\ s,\ m_{\mathrm{K_{DR}}},\ h,\ Ca_i\,]^T$, $t_s$ is the slow dimensionless time variable,
$f_i$ for $i=\{1, \cdots,6\}$, along with $g_1$, and $g_2$ are $O(1)$ functions defined in Appendix \ref{sec:nondim-theta}. System \eqref{eq:slow-theta} is a singularly perturbed problem with fast variables $(V, y_2,\cdots,y_6)$, a slow variable $n$, and a superslow variable $q$, and with small perturbation parameters $\varepsilon_i$ and $\delta$. 
We note that although $V$ and the gating variables $y_i$ evolve on different timescales, they are all faster than $n$ and are therefore grouped together as fast variables. For convenience, we will use the shorthand notation $\varepsilon$ to denote the collection $(\varepsilon_1, \cdots, \varepsilon_6)$ unless specified otherwise. When a numerical value of $\varepsilon$ is specified, it refers to the value of $\varepsilon_2$. 

\begin{remark}\label{re:timescale-separation}
The timescales of system \eqref{eq:slow-theta} can be adjusted via the scaling factors $r_\varepsilon$ and $r_\delta$ (see Table \ref{tab:eps-delta-para}). In particular, increasing $r_\varepsilon$ decreases $\varepsilon_i$ and thus speeds up all fast variables. Increasing $r_\delta$ increases $\delta$ and hence accelerates the superslow variable $q$. By default, we set $r_\varepsilon = r_\delta = 1$, which correspond to the default values of $\varepsilon_i$ and $\delta$ given in Table \ref{tab:eps-delta-para} in Appendix \ref{sec:nondim-theta}.
\end{remark}

\section{Geometric singular perturbation analysis}\label{sec:gspt-theta}

In this section, we apply the extended GSPT framework \cite{Fenichel1979, Nan2015,Vo2013} to the three-timescale cortical theta oscillator model. Our goal is to establish the singular geometric structures that organize the dynamics and identify the mechanisms that can contribute to MMBO dynamics when $0<\varepsilon,\delta\ll1$.

We refer to system \eqref{eq:slow-theta}, which evolves on the slow timescale $t_s$, as the \textit{slow system}.
Equivalent descriptions of the dynamics can be obtained through appropriate time rescaling, resulting in the corresponding \textit{fast system} \eqref{eq:fast-theta} and the \textit{superslow system} \eqref{eq:superslow-theta}, both described in Appendix \ref{ap:singular-limit}. The presence of independent singular perturbation parameters $\varepsilon$ and $\delta$ allows for multiple approaches for implementing GSPT, each leading to a distinct singular-limit prediction. Below, we first provide a brief overview of each singular-limit viewpoint, referring to Appendix \ref{ap:singular-limit} for further details. We then construct singular periodic orbits in the double singular limit $(\varepsilon, \delta) \rightarrow (0,0)$ and illustrate how they perturb to periodic solutions of the full system \eqref{eq:slow-theta} when $\varepsilon$ and $\delta$ are sufficiently small.

\subsection{Singular limit} \label{sub:singular-limit-main}
First, the singular limit $\varepsilon \rightarrow 0$ with $\delta>0$ in the fast system \eqref{eq:fast-theta} leads to the
\textit{fast layer problem} (see \eqref{eq:fast-subsystem-theta}). The set of equilibria of the fast layer problem defines the 2D \textit{critical manifold}, given by   
\begin{equation}\label{eq:ms-theta}
M_{s}:=\{(V, m_{\mathrm {NaP}}, s, m_{\mathrm {K_{DR}}}, h, \mathrm{\mathrm{Ca_i}}, n, q):\ f_1 = f_2 = f_3 = f_4 = f_5 = f_6 = 0\}.
\end{equation}
$M_s$ is divided by its fold curves $L_s$ (saddle-node bifurcations of the fast layer problem; see \eqref{eq:fold-theta}) into three sheets: a lower attracting sheet $M_s^L$, a middle repelling sheet $M_s^M$, and an upper repelling sheet $M_s^U$. By Fenichel's theory \cite{Fenichel1979}, the normally hyperbolic parts of $M_s$ persist, for sufficiently small $\varepsilon$, as locally invariant slow manifolds of the full system \eqref{eq:slow-theta}. On these manifolds, the flow of \eqref{eq:slow-theta} is an $O(\varepsilon)$ perturbation of the reduced slow flow on $M_s$, governed by the 2D \textit{slow reduced problem} \eqref{eq:slow-reduced-theta}, obtained by taking the singular limit $\varepsilon\rightarrow 0$ with $\delta>0$ in the slow system \eqref{eq:slow-theta}. Fenichel theory breaks down near bifurcations of the fast layer problem, including the fold bifurcation curves $L_s$, where normal hyperbolicity is lost.  

To analyze the slow reduced flow on the critical manifold, we derive the \textit{desingularized system} \eqref{eq:desingularization-theta} by projecting the reduced system~\eqref{eq:slow-reduced-theta} onto the $(V,n)$ plane and applying a time rescaling to remove the singularity at the fold curve, as detailed in Appendix \ref{ap:singular-limit}. The desingularized system has two types of singularities: \textit{ordinary singularities} TE, given by \eqref{eq:ordinary-theta}, and \emph{folded singularities} $\mathcal{M}_\delta$, defined by \eqref{eq:folded-singu-orig-theta}. The ordinary singularity is a true equilibrium of the full system \eqref{eq:slow-theta}. In contrast, folded singularities are special points of the slow reduced problem that can allow their trajectories to cross the fold $L_s$ with finite speed. Such solutions are known as singular canards \cite{SW2001, Wechselberger2005, Wechselberger2012}. 
The eigenvalues of the linearization of \eqref{eq:desingularization-theta} at a folded singularity determine its type: a \textit{folded saddle} has real eigenvalues of opposite signs, a \textit{folded node} has real eigenvalues of the same sign, and a \textit{folded focus} has complex eigenvalues. 

In the default parameter regime, the TE is an unstable saddle-focus equilibrium of \eqref{eq:slow-theta} with six negative real eigenvalues and a pair of complex conjugate eigenvalues whose real parts are positive, and the folded singularity is a stable folded node with weak and strong real eigenvalues. The \textit{singular strong canard} of a folded node is the unique trajectory corresponding to its strong stable manifold tangent to the strong eigendirection. The strong canard and the fold curve $L_s$ of $M_s$ form a 2D trapping region, known as the funnel, on the attracting sheet of $M_s$. Trajectories that enter the funnel will be drawn toward the folded node and pass through the fold from an attracting sheet of $M_s$ to a repelling sheet of $M_s$. Such solutions are called singular canards. As $r_\delta$ increases, the stable folded node becomes a folded focus after passing a degenerate folded node point (Fig.~\ref{fig:twopar-hopf-theta}A). Conversely, as $r_\delta$ decreases toward $0$, the folded singularity becomes a folded saddle-node. 

Alternatively, the singular limit $\delta \rightarrow 0$ with fixed $\varepsilon>0$ in the slow system \eqref{eq:slow-theta} yields the 7D \textit{slow layer problem} \eqref{eq:slow-layer-theta}. Its equilibria define the \textit{superslow manifold}:
\begin{equation}\label{eq:mss-theta}
M_{ss}:=\{(V, m_{\mathrm {NaP}}, s, m_{\mathrm {K_{DR}}}, h, \mathrm{Ca_i}, n, q): 
f_1 = f_2 = f_3 = f_4 = f_5 = f_6 = g_1 = 0\},
\end{equation}
which is a 1D subset of the critical manifold $M_s$. Similar to $M_s$, GSPT \cite{Fenichel1979} guarantees that for $\delta$ sufficiently small, the normally hyperbolic subsets of $M_{ss}$ perturb to locally invariant manifolds. On these manifolds, the superslow flow is $O(\delta)$-close to the reduced superslow flow on $M_{ss}$, described by the \textit{superslow reduced problem} \eqref{eq:ssl-reduced-theta}, obtained from taking the singular limit $\delta\rightarrow 0$ with $\varepsilon>0$ in the superslow system \eqref{eq:superslow-theta}. Along $M_{ss}$, the slow layer problem exhibits Hopf bifurcations $M_{ss}^H$. The subsystem Hopf bifurcation $M_{ss}^H$ is also known as a delayed Hopf bifurcation (DHB).

Finally, taking the double singular limit $(\varepsilon, \delta) \rightarrow (0,0)$ in the slow system \eqref{eq:slow-theta} yields the \textit{slow reduced layer problem} \eqref{eq:slow-reduced-layer-theta}. The corresponding desingularized system is given by \eqref{eq:desingularized-double-limit-theta}, which can be obtained as the $\delta\to 0$ limit of the desingularized system \eqref{eq:desingularization-theta} for $\delta>0$. 
In the double singular limit, the ordinary singularities of \eqref{eq:desingularized-double-limit-theta} relax from true equilibria TE of the full system to the superslow manifold $M_{ss}$, while the folded singularity $\mathcal{M}_0$ becomes a canard-delayed-Hopf (CDH) singularity, defined as the intersection of $M_{ss}$ with the fold curve $L_s$ \cite{Vo2013, Letson2017, Phan2023}, and corresponds to a folded saddle-node. CDH singularities arise naturally in three-timescale systems, often serving as organizing centers for local oscillatory behaviors and enabling the canard and DHB mechanisms to coexist and interact.

\begin{remark}\label{rm:FN-DHB-CDH}\rm
Consistent with prior studies \cite{Letson2017, Vo2013, Phan2023}, our analysis shows that the folded singularity $\mathcal{M}_\delta$ obtained from the $\varepsilon$-singular limit (see \eqref{eq:folded-singu-orig-theta} and Fig.~\ref{fig:twopar-hopf-theta}A) is $O(\delta)$-close to the CDH, while the DHB in the $\delta$-singular limit appears to lie $O(\varepsilon)$-close to the CDH (Fig.~\ref{fig:twopar-hopf-theta}B). 
See Appendix \ref{ap:singular-limit} for more details.
\end{remark}


\subsection{Singular orbit construction}\label{sub:singular-orbit}

\begin{figure}[!htp]
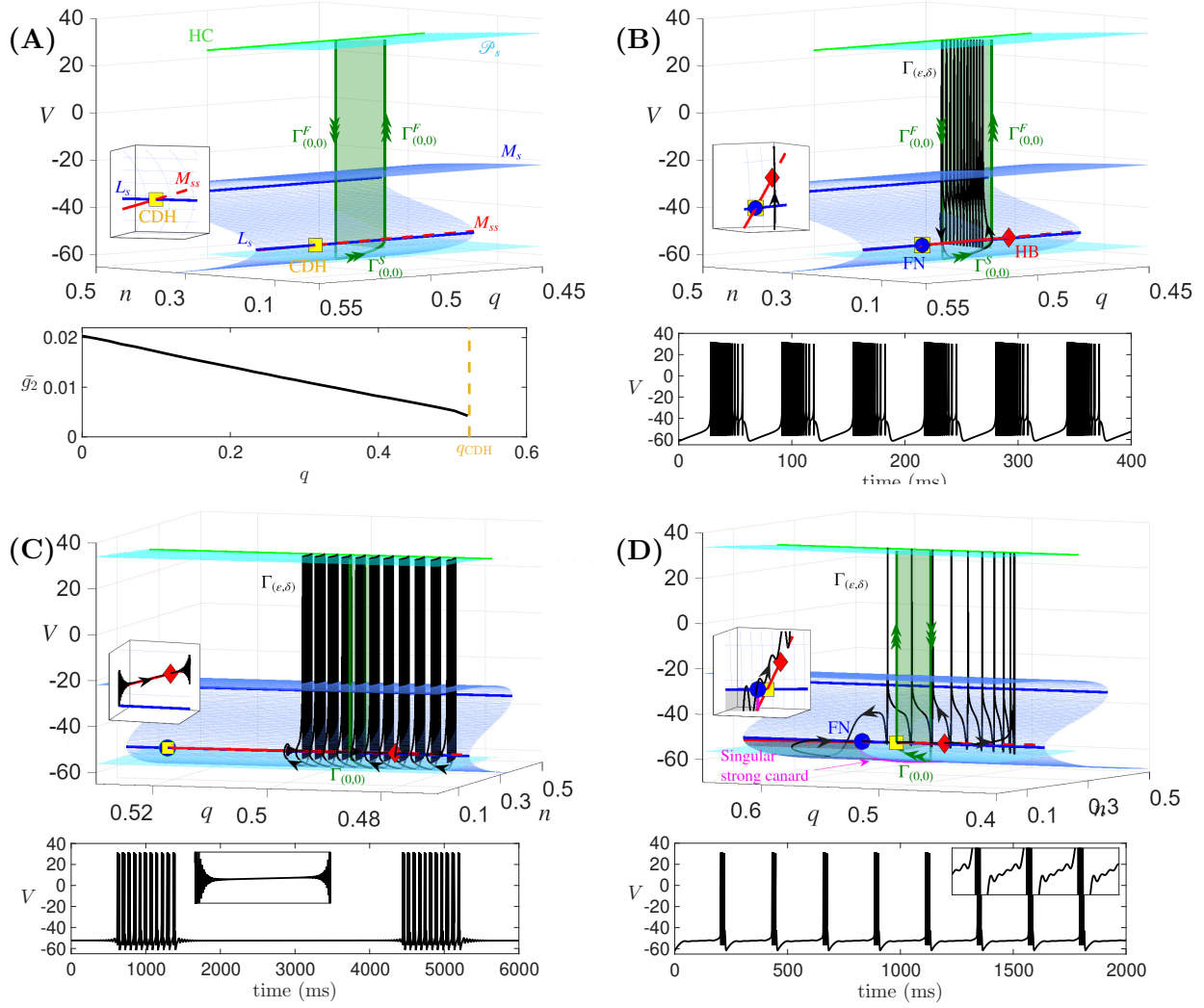

\begin{center}
\begin{tabular}{@{}p{0.48\linewidth}@{\quad}p{0.48\linewidth}@{}}

\begin{minipage}[t][\height][t]{\linewidth}
    \subfigimg[width=\linewidth]{\bfseries{\small{(A)}}}{singular_orbit_for_spiking.eps}
    \vfill   
    \subfigimg[width=0.9\linewidth]{\bfseries{\small{}}}{ave_nullcline_q.eps}
\end{minipage}
&
\begin{minipage}[t][\height][t]{\linewidth}
    \subfigimg[width=\linewidth]{\bfseries{\small{(B)}}}{singular_orbit_and_soln_reps_10_rdel_p00001.eps}  
    \vfill   
    \subfigimg[width=0.9\linewidth]{\bfseries{\small{}}}{time_trace_eps_10_del_p00001_.eps}
\end{minipage}

\\
\begin{minipage}[t][\height][t]{\linewidth}
    \subfigimg[width=\linewidth]{\bfseries{\small{(C)}}}{singular_orbit_and_soln_reps_10_rdel_p01.eps}  
    \vfill   
    \subfigimg[width=0.9\linewidth]{\bfseries{\small{}}}{time_trace_eps_10_del_p01_.eps}
\end{minipage}
&
\begin{minipage}[t][\height][t]{\linewidth}
    \subfigimg[width=\linewidth]{\bfseries{\small{(D)}}}{singular_orbit_for_spiking_at_CDH.eps}
    \vfill   
    \subfigimg[width=0.9\linewidth]{\bfseries{\small{}}}{time_trace_eps_10_del_1_.eps}
\end{minipage}
\end{tabular}
\end{center}
 \caption{Projections of representative singular orbits $\Gamma_{(0,0)}$ (dark green trajectories) onto $(n,q,V)$-space, with (A-C) $q=0.490917<q_\mathrm{CDH}$ and (D) $q=q_\mathrm{CDH}$. The lower panel of (A) plots the function $\bar{g}_2(q)$ over the $0\leq q<q_{\rm CDH}$, showing \eqref{eq:averaged-super-slow} has no equilibrium. Panels (B)-(D) show the perturbed bursting solutions $\Gamma_{(\varepsilon,\delta)}$ (black trajectories) with different perturbation sizes: the upper panels show their projections onto $(n,q,V)$-space, and the lower panels show the voltage traces in time. The parameter values are (B) $(\varepsilon,\delta)=(0.03, 6 \times 10^{-7})$ with $(r_\varepsilon, r_\delta)=(10, 10^{-5})$, (C) $(\varepsilon,\delta)=(0.03, 6 \times 10^{-4})$ with $(r_\varepsilon, r_\delta)=(10, 0.01)$, and (D) $(\varepsilon,\delta)=(0.031, 0.06)$ with $(r_\varepsilon, r_\delta)=(10.397, 1)$. Also shown are the critical manifold $M_s$ (blue surface) and the superslow manifold $M_{ss}$ (red curve). $M_s$ is divided into three sheets by the fold curves $L_s$ (blue curves): the attracting lower sheet $M_s^L$ and the repelling middle and upper sheets ($M_s^M$ and $M_s^U$). $M_{ss}$ consists of an attracting branch $M_{ss}^a$ (solid) and a repelling branch $M_{ss}^r$ (dashed). The lower fold $L_s$ intersects $M_{ss}$ at the CDH singularity (yellow square). The cyan surfaces represent the periodic attractor $\mathcal{P}_s$, which emanates from a curve of supercritical Hopf bifurcation on $M_s^U$ (not shown) and terminates at the homoclinic bifurcation (HC, light green curve). The DHB and the folded node (FN) are denoted by the red diamond and blue circle, respectively. In panel (D), the trajectory enters the funnel (shaded region) of the folded node, which is bounded by the lower fold curve $L_s$ and the singular strong canard (magenta curve). Note that the displayed $q$ range varies across panels.}
 \label{fig:singular-orbit}
\end{figure}

In this subsection, we construct singular periodic orbits $\Gamma_{(0,0)}$ of \eqref{eq:slow-theta} in the double singular limit $(\varepsilon, \delta) \rightarrow (0,0)$ by concatenating solution segments of singular limit systems described above in Subsection \ref{sub:singular-limit-main}. We then use the singular orbits to understand the full-system trajectories for $0<\varepsilon,\delta\ll 1$. 
We denote the fast, slow, and superslow flows, i.e., solutions of the fast layer problem \eqref{eq:fast-subsystem-theta}, the slow reduced layer problem \eqref{eq:slow-reduced-theta}, and the superslow reduced problem \eqref{eq:ssl-reduced-theta}, by $\Gamma^F_{(0,0)}$, $\Gamma^S_{(0,0)}$, and $\Gamma^{SS}_{(0,0)}$, respectively. According to GSPT, the singular orbits $\Gamma^F_{(0,0)}\cup \Gamma^S_{(0,0)} \cup \Gamma^{SS}_{(0,0)}$ perturb to periodic solutions $\Gamma_{(\varepsilon, \delta)}$ of the full system for sufficiently small perturbations $0< \varepsilon, \delta \ll 1$. 

Fig. \ref{fig:singular-orbit} shows the projection of the critical manifold $M_s$ (blue surface) and the superslow manifold $M_{ss}$ (red curve) onto the $(n,q,V)$-space. $M_s$ is separated by the fold curves $L_s$ (blue curves) into three sheets: $M_s^L$ is attracting, while $M_s^M$ and $M_s^U$ are repelling. The lower fold $L_s$ intersects $M_{ss}$ at the CDH singularity (yellow square), which, as discussed above, coincides with the folded singularity and the DHB in the double singular limit (Fig. \ref{fig:singular-orbit}A, top panel). 
In this case, the CDH separates the attracting branch $M_{ss}^a$ (solid red curve) from the repelling branch $M_{ss}^r$ (dashed red curve). The maximum and minimum values of $V$ along the periodic orbit attractor $\mathcal{P}_s$ of the fast layer problem \eqref{eq:fast-subsystem-theta} are indicated by the cyan surface. $\mathcal{P}_s$ emerges via a curve of supercritical Hopf bifurcations along the upper sheet $M_s^U$ (not shown) and terminates along a curve of homoclinic (HC) bifurcations (green curve). 

For the construction of transient singular orbits, we can choose an initial condition on the attracting branch of the superslow manifold $M_{ss}^a$. Transients $\Gamma^{SS}_{(0,0)}$ evolve along $M_{ss}^a$ in the decreasing $q$ direction, governed by the superslow reduced problem \eqref{eq:ssl-reduced-theta}, until they encounter the CDH point. At this point, the singular orbit transitions to a bursting solution $\Gamma^F_{(0,0)}\cup \Gamma^S_{(0,0)}$ with fixed $q=q_{\rm CDH}$, arising from alternation between slow evolution along the lower attracting sheet of $M_s$ and fast spiking phase that initiates at the lower fold $L_s$ and terminates at the HC. Moreover, for each fixed $0\leq q\leq q_\mathrm{CDH}$, the singular attractor is a bursting orbit with that value of $q$. To illustrate this, panel (A) shows one such singular orbit with fixed $q=0.490917<q_{\rm CDH}$, which consists of slow flow $\Gamma^S_{(0,0)}$ (green trajectory with double arrows) and fast spiking along $\mathcal{P}_s$. For clarity, only the first and the last spikes are shown (triple green arrows), while the other spikes are represented by the green surface, indicating a continuum of spiking activity. 
Thus, in the double singular limit, there exists a continuum of bursting singular orbits for $0 \leq q \leq q_{\mathrm{CDH}}$, each corresponding to a fixed value of $q$.
  
To determine whether there exists a stable attractor within this continuum of singular orbits to which the full-system trajectory $\Gamma_{(\varepsilon,\delta)}$ converges in the singular limit $(\varepsilon,\delta)\to (0,0)$, we use the averaging method \cite{sanders2007averaging,Vo2013,WR2017} to derive an averaged superslow reduced problem:
\begin{equation}\label{eq:averaged-super-slow}
    \begin{array}{rcl}
        \frac{dq}{dt_{ss}} &=& \frac{1}{T(q)} \int_{0}^{T(q)}  g_2(Ca_{\mathrm{i}(0, 0)}(\tilde{t},q),q) d\tilde{t} \equiv \bar{g}_2(q).
    \end{array}
\end{equation}
Here, $Ca_{\mathrm{i}(0, 0)}(\tilde{t},q)$ denotes the $Ca_{\rm i}$-coordinate of the singular orbit attractor with period $T(q)$ for a fixed $q$ value. A stable equilibrium of \eqref{eq:averaged-super-slow}, at which there is no net drift in $q$, corresponds to a stable singular orbit attractor of the full system. The bottom panel of Fig. \ref{fig:singular-orbit}A shows the averaged vector field $\bar{g}_2(q)$ over the $0\leq q<q_{\rm CDH}$ interval. Since $\bar{g}_2(q)$ remains positive, the averaged equation \eqref{eq:averaged-super-slow} has no equilibrium; instead, $q$ always drifts in the increasing direction.  
 
With the construction of singular orbits $\Gamma_{(0,0)}$ established, we next illustrate how these orbits perturb to different full-system trajectories $\Gamma_{(\varepsilon, \delta)}$ (black curves) as $\delta$ increases from $0$ to $O(0.1)$, while $\varepsilon$ is held fixed at $O(0.01)$ (Fig.~\ref{fig:singular-orbit}B-D). 
In each panel, the voltage trace of $\Gamma_{(\varepsilon, \delta)}$ is shown below its $(n,q,V)$-projection. 
For nonzero $\varepsilon$ and $\delta$ values, the DHB (red diamond) and the folded singularity (blue circle) no longer coincide with the CDH, but instead lie a small distance away from it (see Remark \ref{rm:FN-DHB-CDH}). 
The folded singularity is a folded node (denoted as FN). As $\delta$ increases from panel (B) to panel (D), the FN moves farther from the CDH point toward larger $q$ values, while the CDH position remains unchanged. The DHB lies $O(\varepsilon)$ away from the CDH and separates the attracting and repelling branches of the superslow manifold $M_{ss}$. 
The full-system equilibrium lies outside the displayed window and does not influence the dynamics of \eqref{eq:slow-theta}.

Fig.~\ref{fig:singular-orbit}B shows $\Gamma_{(\varepsilon, \delta)}$ with $(\varepsilon,\delta) = (O(0.01), O(10^{-6}))$ is a fold–HC bursting trajectory with $q$ effectively frozen. As $\delta$ increases further, $\Gamma_{(\varepsilon, \delta)}$ exhibits MMBO dynamics (Figs.~\ref{fig:singular-orbit}C and D). Moreover, the number of bursting events within each periodic MMBO cycle decreases as $\delta$ increases, and the mechanism underlying the SAOs switches from DHB to CDH. 
Specifically, Fig.~\ref{fig:singular-orbit}C shows that $\Gamma_{(\varepsilon,\delta)}$ with $(\varepsilon,\delta) = (O(0.01), O(10^{-3})$ undergoes a sequence of bursting episodes, during which there is a superslow net increase in $q$. This behavior reflects the existence of a continuum of bursting singular orbits for $q\leq q_{\rm CDH}$ and that the averaged vector field of $q$ in \eqref{eq:averaged-super-slow} is positive. Each burst is initiated by a fast jump at a regular fold point and terminates at the HC bifurcation, as the bursting orbit in panel (B).  
As the trajectory passes the DHB, bistability arises between $M_{ss}^a$ and the bursting attractors. However, since $M_{ss}^a$ lies above the fold curve $L_s$, the trajectory on the lower branch of $M_s$ crosses the fold first to produce additional bursts, rather than approaching $M_{ss}^a$. As the trajectory gets closer to the CDH point where $M_{ss}^a$ and $L_s$ intersect, it is eventually attracted to $M_{ss}^a$, thereby terminating the bursting sequence.  
Afterward, the trajectory follows $M_{ss}^a$ on the superslow timescale in the direction of decreasing $q$, generating SAOs with decreasing amplitudes as it approaches the DHB. After passing the DHB, the trajectory undergoes a superslow drift along the repelling branch $M_{ss}^r$, during which SAOs occur with increasing amplitudes. When the orbit has moved sufficiently far from the DHB, it peels off $M_{ss}^r$ and initiates a fast jump to large $V$, thereby entering the bursting phase and completing the cycle. In this regime, the orbit remains near the DHB, with the CDH and FN points relatively far away, so the SAOs are organized solely by the DHB mechanism. 

As $\delta$ increases further to $O(0.1)$ (Fig.~\ref{fig:singular-orbit}D), the MMBO trajectory $\Gamma_{(\varepsilon,\delta)}$ consists of only a single bursting event per cycle, followed by SAOs during the silent phase. After passing the DHB, the geometric deconstruction of the periodic MMBO in panel (D) is similar to that in panel (C). The main difference is that the larger value of $\delta$ in (D) allows for faster accumulation of $q$, bringing the orbit into the funnel of the FN after a single burst, where the funnel is the shaded region in $M_s^L$ bounded by the fold curve $L_s$ and the singular strong canard (magenta curve). As a result, the orbit is guided toward the FN along $M_{ss}^a$ and subsequently passes through the CDH and DHB, which completes the cycle. In this case, the SAOs are organized by both the canard dynamics and the DHB, reflecting the interaction between the two mechanisms near the CDH singularity. 

Thus, we have shown that the singular orbits $\Gamma_{(0,0)}$ provide faithful predictions for the fully perturbed trajectories $\Gamma_{(\varepsilon,\delta)}$ of \eqref{eq:slow-theta} when $0<\varepsilon \ll 1$ and $0<\delta \ll 1$. For simplicity, in later geometric descriptions of $\Gamma_{(\varepsilon,\delta)}$, we omit the singular orbits and directly refer to the different segments of $\Gamma_{(\varepsilon,\delta)}$ as being governed by the GSPT-derived subsystems.
Within this GSPT regime, $\varepsilon$ primarily controls the fast-slow timescale separation and the $O(\varepsilon)$ distance between the DHB and CDH, whereas $\delta$ controls the superslow drift of $q$ and the $O(\delta)$ separation between the FN and CDH. For fixed $\varepsilon$, very small $\delta$ keeps $q$ effectively frozen, so the trajectory shadows a fixed-$q$ bursting singular orbit. Increasing $\delta$ allows $q$ to drift sufficiently after either multiple bursting episodes or a single burst for the trajectory to approach the attracting branch of the superslow manifold $M_{ss}^a$ and generate MMBOs with SAOs. Although increasing $\delta$ moves the FN farther away from the CDH, it also increases the amount of $q$-drift during bursting, allowing the trajectory to enter the funnel of the FN. As a result, the SAO-generating mechanism shifts from being purely DHB-driven to being organized by the CDH, where folded-node canard and DHB mechanisms interact. 
 

\begin{remark}
The analysis and trajectories shown in Fig.~\ref{fig:singular-orbit} focus on the effect of increasing $\delta$ for fixed small $\varepsilon$. While we do not show the trajectories obtained by increasing $\varepsilon$ for fixed $\delta$, the expected effects can be inferred from the GSPT analysis discussed above. In particular, increasing $\varepsilon$ slows the fast variables and moves the DHB farther from the CDH and FN. The first effect reduces the number of spikes within each bursting episode, so we expect fewer spikes per burst as $\varepsilon$ increases. The second effect separates the DHB and canard mechanisms. Thus, when SAOs are primarily DHB-mediated, as in Fig.~\ref{fig:singular-orbit}C, their generating mechanism is expected to remain DHB-mediated, whereas for MMBOs that are organized by the CDH, such as the trajectory in Fig.~\ref{fig:singular-orbit}D, increasing $\varepsilon$ is expected to shift the SAOs toward a more DHB-mediated mechanism. We examine these effects of $\varepsilon$ in more detail in Sections~\ref{sec:numerical-simu-theta} and~\ref{sec:mmos-mmbos-transition}.
\end{remark}

\section{Full-system bifurcation analysis}\label{sec:numerical-simu-theta}

The GSPT analysis in Section \ref{sec:gspt-theta} identifies geometric mechanisms for MMBOs and provides predictions of the full-system dynamics for $0<\varepsilon, \delta\ll 1$. In this section, we perform a numerical continuation study of the full system \eqref{eq:slow-theta} and investigate how the periodic orbits of the full system are organized for broader ranges of $\varepsilon, \delta > 0$, including regimes where the singular orbit predictions become less accurate or break down.

Below, we first investigate the effects of increasing $r_\varepsilon$ (i.e, decreasing $\varepsilon$, see Remark \ref{re:timescale-separation}) on the dynamics of the full system \eqref{eq:slow-theta} in Section \ref{sub:effect-fast-timescales}, where we perform a full-system bifurcation analysis by treating $r_\varepsilon$ as the bifurcation parameter. 
We then perform a two-parameter study in Section \ref{sub:effect-solution-types}, where several key bifurcations identified from the one-parameter bifurcation analysis are continued in the $(r_\varepsilon, r_\delta)$-plane. The bifurcation analysis was performed using the numerical continuation package AUTO  \cite{doedel1981auto, doedel1997auto97}.

\subsection{One-parameter bifurcation analysis of the full system} 
\label{sub:effect-fast-timescales}

\begin{figure}[!htp]
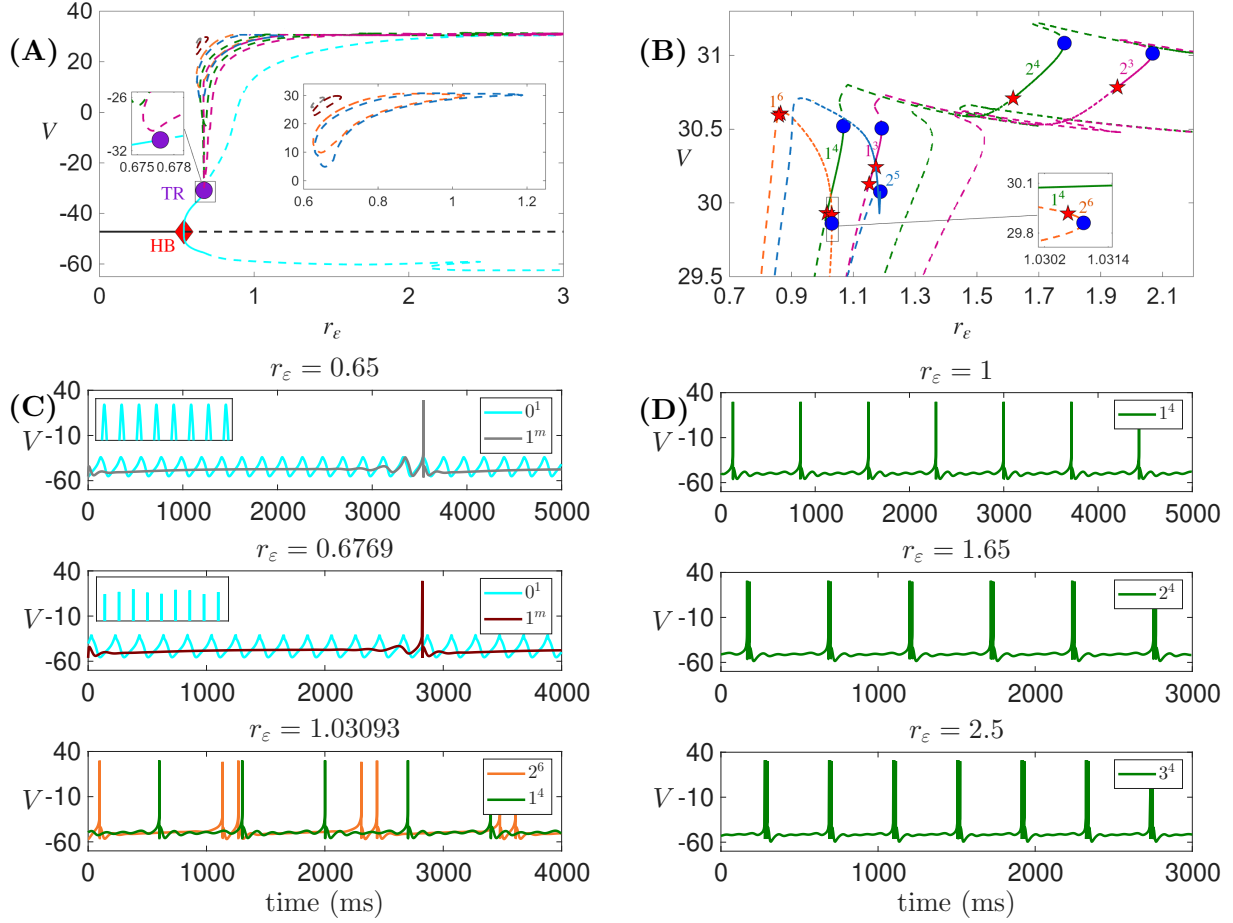

\begin{center}
\begin{tabular}{@{}p{0.48\linewidth}@{\quad}p{0.48\linewidth}@{}}
\subfigimg[width=\linewidth]{\bfseries{\small{(A)}}}{bd_V_r.eps}&\subfigimg[width=\linewidth]{\bfseries{\small{(B)}}}{bd_V_r_zoomin.eps}
\\
\subfigimg[width=0.95\linewidth]{\makebox[0pt][l]{\raisebox{1.2ex}[0pt][0pt]{\bfseries{\small{(C)}}}}}{timetrace_ic1.eps}&\subfigimg[width=0.95\linewidth]{\makebox[0pt][l]{\raisebox{1.2ex}[0pt][0pt]{\bfseries{\small{(D)}}}}}{timetrace_ic2.eps}
\end{tabular}
\end{center}
 \caption{Effects of variations in the fast timescales $r_\varepsilon$ on the dynamics of \eqref{eq:slow-theta}. (A) One-parameter bifurcation diagram of the full system \eqref{eq:slow-theta} with respect to $r_\varepsilon$ for $r_\delta=1$. Other parameters are fixed at their default values in Table \ref{tab:para-theta}. The solid (resp., dashed) black curve denotes stable (resp., unstable) full-system equilibria. The red diamond denotes the full-system Hopf bifurcation (HB), giving rise to a periodic orbit branch (cyan). This branch changes stability from stable (solid) to unstable (dashed) at a torus bifurcation (TR, purple circle). Isolas of MMO- and MMBO-type periodic oscillations are shown above the cyan branch (solid: stable, dashed: unstable). The inset shows the enlarged views of the gray, brown, orange, and light blue isolas.
 (B) Enlarged view of the isolas corresponding to MMBO orbits with $6$-SAOs (orange), $5$-SAOs (light blue), $4$-SAOs (green), and $3$-SAOs (pink). Along each isola, stable segments (solid curves) lie between a period-doubling bifurcation (PD, red star) and a saddle-node bifurcation of periodic orbits (SNPO, blue circle). (C)-(D) Example voltage traces of \eqref{eq:slow-theta} for different values of $r_\varepsilon$. } 
 \label{fig:bd}
\end{figure}

The bifurcation diagrams of the full system \eqref{eq:slow-theta} are presented in Fig. \ref{fig:bd}A with $r_{\varepsilon}$ as the bifurcation parameter while all other parameters are fixed at their default values. Recall that increasing $r_\varepsilon$ corresponds to speeding up all fast variables. 

The full-system equilibrium at the singular limit of $\varepsilon=0$ persists for $\varepsilon>0$ when $\delta$ is fixed at a small nonzero value. As $\varepsilon$ increases, equivalently as $r_\varepsilon$ decreases, this equilibrium becomes stable in a Hopf bifurcation (HB, red diamond) at $r_{\varepsilon,\mathrm {HB}}\approx 0.5458$. In Fig.~\ref{fig:bd}A, $\varepsilon=0$ corresponds to the limiting direction $r_\varepsilon\to\infty$, to the right of the displayed range.
The HB is supercritical and gives rise to a stable branch of periodic orbits (solid cyan) that becomes unstable (dashed cyan) in a torus (TR) bifurcation (purple circle) at $r_{\varepsilon,\mathrm {TR}} \approx 0.67689$. In addition, other branches of periodic solutions of the full system emerge as isolated closed curves in this parameter space, commonly referred to as \textit{isolas}, which are often observed in fast-slow dynamical systems exhibiting MMO-, bursting-, or MMBO-type complex oscillations \cite{fernandez1997isolas,koksal2020canard, tsaneva2010full, barrio2020spike, Farjami2020, vo2012bifurcations,vo2026symmetric, barrio2024dynamics}. 
Despite the presence of many isolas, we show only a representative selection of this family (see the colored curves above the cyan branch in panel (A); also see panel (B) for an enlarged view). Example stable solutions from these periodic orbit branches are shown in panels (C) and (D), with the same color coding as the corresponding isola branches. We use the notation $L^s$ to denote SAO-LAO rhythms, where $L$ denotes the number of LAOs and $s$ the number of SAOs per cycle. We refer to these rhythms as MMOs when $L=1$ and as MMBOs when $L>1$ (with $s>0$), although MMOs can be viewed as simple cases of MMBOs. 
Along each of these isolas, solutions have a fixed number of SAOs, while the number of LAOs varies. Varying $r_\varepsilon$ can move an orbit from one isola to another, thereby also changing the number of SAOs. 

Between the HB and TR bifurcations, the full system exhibits stable SAO periodic orbits. Near the TR point, there exist small regions of bistability between the cyan SAO branch and isolas of $1^m$ MMO orbits. For example, bistability occurs between the cyan branch and the gray isola (see example solutions in Fig.~\ref{fig:bd}C, top row), and between the cyan branch and the brown isola (see example solutions in Fig.~\ref{fig:bd}C, middle row). For convenience, we treat the TR bifurcation as an approximate transition from purely subthreshold SAOs to MMOs, although it does not precisely mark the onset of MMOs due to the existence of bistability. On the gray and brown isolas (Fig.~\ref{fig:bd}A, inset), stable $1^m$ solutions lie on top between two PD bifurcations, which are not labeled. Solutions along these MMO isolas retain a single full spike: although the SAO preceding the LAO changes in amplitude along each isola, it never develops into a full spike. 
A more detailed examination of the solution behaviors along MMO isolas of \eqref{eq:slow-theta} is left for future work.

Next, we examine how MMBOs are organized along their isolas, a subset of which is shown in Fig. \ref{fig:bd}B. The complete isola curves corresponding to 6 (orange) and 5 SAOs (light blue) are also shown in the inset of Fig.~\ref{fig:bd}A.  An MMBO isola contains multiple stable segments, with each segment bounded by a period-doubling bifurcation (PD, red star) and a saddle-node bifurcation of periodic orbits (SNPO, blue circle). For example, the orange isola associated with the 6-SAO orbits contains two stable segments with $1^6$ and $2^6$ orbit attractors (see Fig.~\ref{fig:bd}B). 
For clarity, solid curves are used only for stable segments that correspond to solutions with fully formed spikes; other stable segments are not distinguished from unstable ones (dashed curves).
Overlap between stable segments on different isolas gives rise to multistability in parameter space; see the inset in panel (B) for an example of bistability between the green and orange isolas, with the corresponding stable orbits shown in the bottom row of panel (C). Together, Figs.~\ref{fig:bd}A and B show that isolas closer to the TR bifurcation have smaller widths and exhibit more SAOs, whereas those farther away contain more stable segments and a greater number of full spikes (or LAOs). 

To illustrate the route from purely subthreshold SAOs to MMOs, and subsequently to MMBOs via a spike-adding cascade, we focus on the green isola formed by the 4-SAO orbits, along which the solution changes from $1^4$ to $2^4$ to $3^4$ and so forth (see Fig.~\ref{fig:bd}D for representative voltage traces). For each $L=0, 1, 2, \cdots$, there exists a small interval of parameter values $r_\varepsilon$ along this isola over which the full system exhibits a continuous transition from periodic solutions with $L$ large spikes to those with $L+1$ spikes, while maintaining four SAOs. Starting from $0^4$ subthreshold SAO solutions, increasing $r_\varepsilon$ leads to the emergence of a stable $1^4$ orbit after a PD bifurcation. This branch loses stability at an SNPO bifurcation, initiating the formation of the second spike. The spike fully develops at the next PD bifurcation, yielding a stable $2^4$ orbit branch with two fully formed spikes. Such spike-adding transition from stable $L^4$ to $(L+1)^4$ MMBOs repeats through successive SNPO and PD bifurcations as $r_\varepsilon$ continues to increase, up to a maximal number of spikes that can be formed along this isola. Through numerical continuation in AUTO \cite{doedel1981auto, doedel1997auto97}, we identify stable solutions with up to 11 spikes (i.e., a $11^4$ segment) for $r_\varepsilon \in (10.3939, 10.3999)$ (see Fig.~\ref{fig:singular-orbit}D); we do not extend the continuation further to determine the exact upper bound. In the limit of $r_\varepsilon\to \infty$ (i.e., in the singular limit $\varepsilon\to 0$) the GSPT analysis in Section \ref{sec:gspt-theta} predicts that the number of spikes goes to infinity (Fig.~\ref{fig:singular-orbit}).  

\subsection{Two-parameter bifurcation analysis of the full system}
\label{sub:effect-solution-types}

\begin{figure}[!htp]
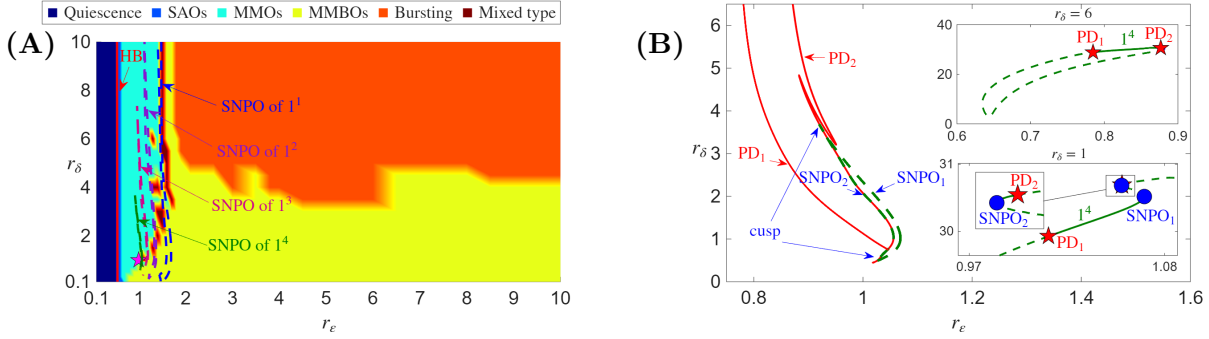

\begin{center}
\begin{tabular}
{@{}p{0.48\linewidth}@{\quad}p{0.48\linewidth}@{}}
\subfigimg[width=\linewidth]{\bfseries{\small{(A)}}}{classify_all_2026_v1.eps} &
\subfigimg[width=\linewidth]{\bfseries{\small{(B)}}}{2par_bif_isolas_4SAOs.eps}
\end{tabular}
\end{center}
\caption{Effects of variations in the fast and superslow timescales, $r_\varepsilon$ and $r_\delta$, on the dynamics of \eqref{eq:slow-theta}. (A) As $r_\varepsilon$ and $r_\delta$ are varied over the ranges $0.1 \leq r_\varepsilon \leq 10$ and $0.1 \leq r_\delta \leq 10$, the full system exhibits different activity patterns: quiescence (dark blue region), SAOs without spiking (light blue region), MMOs (cyan region), MMBOs (yellow region), bursting (orange region), and mixed MMO-MMBO type (red region). Also shown are the full-system Hopf bifurcation curve (red curve) and four SNPO curves corresponding to $1^s$ MMOs with $s \geq 1$ (dashed dark green, pink, purple, and blue curves). The magenta star marks MMOs obtained at the default timescales (Fig. \ref{fig:bd}D, top panel). (B) Enlarged view of the two-parameter bifurcation curves of the SNPO (dashed green, same curve as in panel (A)) and PD (solid red) associated with the stable $1^4$ orbit segment. Partial isolas for fixed $r_\delta=1$ and $r_\delta=6$ are shown in the insets.
} 
\label{fig:all-transitions}
\end{figure} 


Fig.~\ref{fig:all-transitions}A summarizes the effects of varying the fast and superslow timescales on the dynamics of the full system \eqref{eq:slow-theta}. The results are obtained by numerically simulating the full system on a grid in $(r_\varepsilon, r_\delta) \in [0.1,10]\times [0.1,10]$ and classifying the final solution states as quiescence (dark blue), SAOs (light blue), MMOs ($1^s$, cyan), MMBOs ($L^s$ with $L>1$, yellow), regular bursting (orange), and mixed MMO/MMBO types (dark red). Because of the complexity of isola structures (Fig.~\ref{fig:bd}), different initial conditions may lead to different activity patterns within the same parameter region. We therefore do not attempt to distinguish these regions. 

We compute, in the $(r_\varepsilon, r_\delta)$ parameter plane, the locations of several key bifurcations identified in the one-parameter bifurcation diagram. The solid red curve denotes the full-system Hopf bifurcation (red diamond in Fig.~\ref{fig:bd}A), which separates quiescence from SAO dynamics. 
To characterize the transition from MMO to MMBO dynamics, we compute the SNPO bifurcations on stable $1^s$ MMO branches for $s=\{1, 2, 3, 4\}$ (green, pink, purple and blue dashed curves in Fig.~\ref{fig:all-transitions}A). For a fixed $s$, such an SNPO marks where the corresponding stable $1^s$ MMO branch loses stability. However, this bifurcation should not be interpreted as an exact transition between MMO and MMBO dynamics as discussed before: increasing $r_{\varepsilon}$ past the SNPO of one stable MMO branch does not typically carry the trajectory to the next stable MMBO branch. Instead, the trajectory may jump to a nearby stable MMO segment that lies on a different isola. Thus, crossing an SNPO curve in Fig.~\ref{fig:all-transitions}A only indicates the loss of stability of a particular $1^s$ MMO branch, but not necessarily the transition from MMOs to MMBOs. Nonetheless, the SNPO curve of the $1^1$ branch (Fig.~\ref{fig:all-transitions}A, blue dashed curve) provides a useful approximate upper bound of $r_\varepsilon$ for the MMO region, since all stable $1^s$ segments with $s>1$ lie to its left. As a result, increasing $r_\varepsilon$ beyond the $1^1$ SNPO curve moves the system into a regime where no stable MMO solutions remain. Fig.~\ref{fig:all-transitions}A shows that for $r_\delta$ relatively small, the blue SNPO curve lies outside the MMO region, whereas for relatively large $r_\delta$, it closely aligns with the MMO/MMBO boundary. Our numerical simulations suggest that the transition from MMOs to MMBOs is sometimes accompanied by irregular alternations between blocks $1^{s_1}$ and $2^{s_2}$ (denoted as mixed type, orange region) due to the overlapping of isolas. Similar chaotic alternations between $n$- and $n+1$-spike periodic orbits have been reported in other studies \cite{tsaneva2010full, barrio2020spike}. 

To understand why the blue SNPO curve changes from a loose upper bound to a close approximation of the MMO/MMBO boundary, we next examine how the SNPO curves are organized in the two-parameter plane and how the corresponding isola structures evolve as $r_\delta$ increases.
As a representative example, we focus on the isolas with 4 SAOs and the associated SNPO bifurcations. 
Fig.~\ref{fig:all-transitions}B shows an enlarged view of the $1^4$ SNPO curve (green) together with the corresponding $1^4$ PD bifurcation curve (red). The lower inset displays a subset of the 4-SAO isola for $r_\delta=1$, projected onto the $(r_\varepsilon, V)$ plane (i.e., the green isola in Fig.~\ref{fig:bd}B).\footnote{The stable segment bounded by SNPO$_2$ and PD$_2$ in Fig.~\ref{fig:all-transitions}B inset corresponds to stable $1.5^4$ orbits and was not indicated in Fig.~\ref{fig:bd}B for simplicity.} The bifurcations SNPO$_1$ and SNPO$_2$ (blue circles) correspond to the right and left branches of the SNPO curve in the $(r_\varepsilon,r_\delta)$ plane. 
Starting at $r_\delta=1$, the isola contains multiple stable $L^4$ segments through the $r_\varepsilon$-induced spike-adding cascade as discussed before. As $r_\delta$ increases, the stable segments with $L>2$ gradually disappear, ultimately leaving only two stable segments: $1^4$ bounded by PD$_1$ and SNPO$_1$, and $1.5^4$ bounded by SNPO$_2$ and PD$_2$, where the additional 0.5 corresponds to a larger-amplitude subthreshold excursion. After passing the upper SNPO cusp bifurcation, these stable segments coalesce and become a stable $1^4$ segment bounded by PD$_1$ and PD$_2$ 
(see the upper inset in Fig.~\ref{fig:all-transitions}B for the isola with $r_\delta=6$). With further increase in $r_\delta$, the PD bifurcations also coalesce and disappear, after which the entire isola vanishes, a scenario commonly associated with the creation and destruction of isolas \cite{fernandez1997isolas}. 

Thus, increasing $r_\delta$ progressively simplifies the isola structure by gradually eliminating higher-$L$ stable MMBO segments, and eventually destroying the entire isola. As shown in Fig.~\ref{fig:all-transitions}A, the SNPO curves associated with $1^s$ orbits disappear successively as $r_\delta$ increases, with larger-$s$ curves vanishing first. 
Consequently, for large $r_\delta$, any remaining isolas with $s>1$ are simpler than their small-$r_\delta$ counterparts: they contain fewer stable MMBO segments, or only a single stable MMO segment. As a result, crossing a $1^s$ SNPO curve with $s>1$ becomes less relevant for MMO-to-MMBO transitions, and the $1^1$ SNPO curve provides an almost exact MMO/MMBO transition boundary. For small $r_\delta$, however, the $1^1$ SNPO curve serves only as a loose upper bound because of the persistence of additional $1^s$ SNPO curves and the greater complexity of the associated isola structures.

Finally, we describe the transition from MMBOs to regular bursting in Fig.~\ref{fig:all-transitions}A.
For small $r_\delta$, MMBOs are robust to increasing $r_\varepsilon$ (Fig.~\ref{fig:all-transitions}A, lower yellow region), whereas for larger $r_\delta$, the MMBO regime becomes highly sensitive to increases in $r_\varepsilon$ and quickly transitions to the regular bursting dynamics. Similar to the mixed-type solutions observed near the MMO/MMBO transition, we also observe the occasional appearance of an MMBO block within the bursting solutions. For simplicity, we classify these solutions as regular bursting. 
\section{Transitions between different activity patterns} \label{sec:mmos-mmbos-transition}


The singular analysis in Section \ref{sec:gspt-theta} predicts how the two timescale parameters control the MMBO mechanisms: $\varepsilon$ controls the fast-slow separation and the distance between the DHB and CDH, whereas $\delta$ controls the drift of the superslow variable and the distance between the folded node and CDH. The full-system bifurcation analysis in Section \ref{sec:numerical-simu-theta} shows how the fully-perturbed periodic orbits are organized along isolas over a broad range of $\varepsilon, \delta>0$. It also identifies various transitions between MMOs, MMBOs, and regular bursting in the $(\varepsilon,\delta)$ parameter space, as well as the persistence of MMO and MMBO dynamics outside the singular-limit regime where the GSPT analysis no longer applies. In this section, we combine these results to determine how changes in timescale separation drive these transitions.

We begin by analyzing the MMO dynamics at the default timescale parameter values (i.e., $r_\varepsilon=r_\delta=1$) in subsection \ref{sub:mmos-default}, which provides the reference case for the subsequent transitions. 
We then show in subsection \ref{sub:exaggeration-theta} that strengthening the three-timescale separation can produce transitions from MMOs to MMBOs by two routes: decreasing $\varepsilon$ changes the underlying SAO mechanism from a DHB-dominated mechanism to a CDH mechanism, whereas decreasing $\delta$ preserves a DHB-dominated mechanism. We next examine why MMOs and MMBOs persist when the system is moved away from the singular limit in subsection \ref{sub:mmos-preservation-theta}, and finally describe the two mechanisms by which MMBOs lose their SAOs and transition to regular bursting in subsection \ref{sub:mmbos-bursting}.


\subsection{Analysis of MMOs with the default timescales}\label{sub:mmos-default}

\begin{figure}[!t]
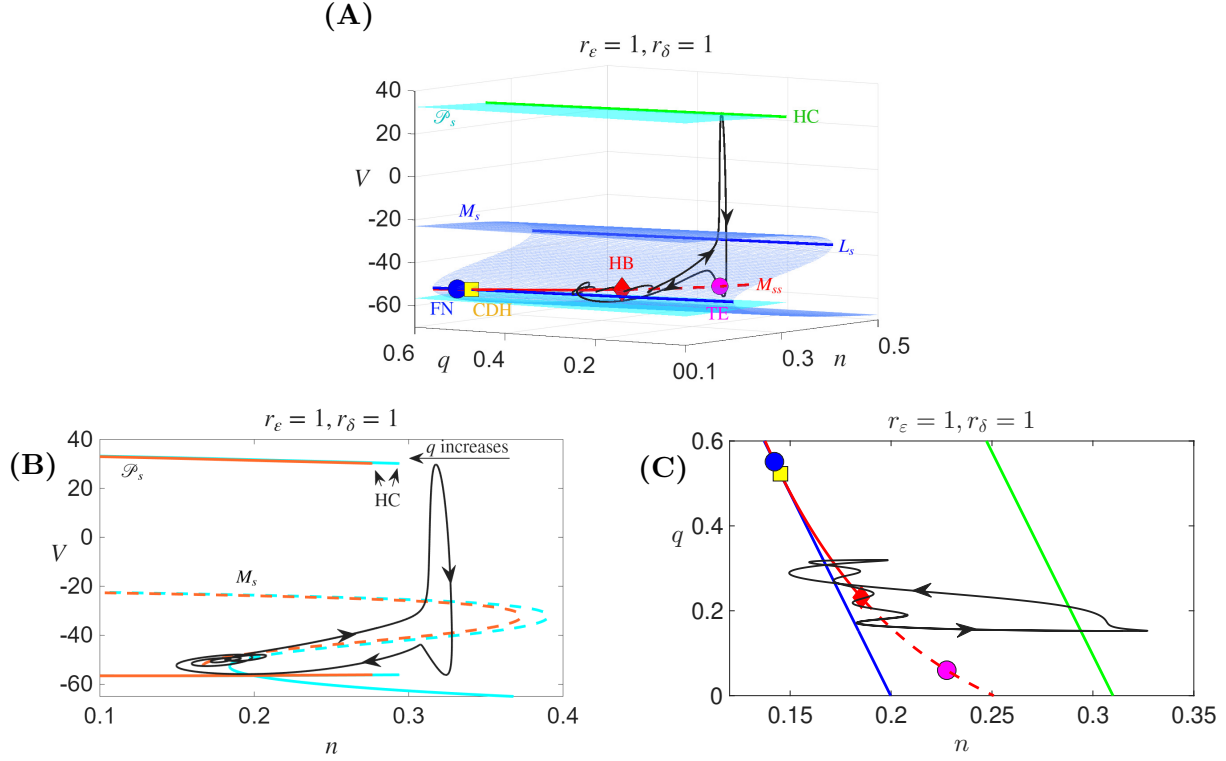

\begin{center}
\begin{tabular}
{@{}p{0.48\linewidth}@{\quad}p{0.48\linewidth}@{}}
\multicolumn{2}{c}{\subfigimg[width=0.48\linewidth]{\bfseries{\small{(A)}}}{default_projection_nqv.eps}}\\
\subfigimg[width=\linewidth]
{\bfseries{\small{(B)}}}{fastsys_bif.eps}&
\subfigimg[width=\linewidth]{\bfseries{\small{(C)}}}{twopar_hc_nq.eps}
\end{tabular}
\end{center}
\caption{Different projections of an attracting MMO solution of the full-system \eqref{eq:slow-theta} for $(r_\varepsilon,r_\delta)=(1,1)$, with default $(\varepsilon_i, \delta)$ values given in Table  \ref{tab:eps-delta-para}. (A) In $(n,q,V)$-space, the critical manifold $M_s$ (blue surface), two fold curves $L_s$ (blue curves), the superslow manifold $M_{ss}$ (red curve), the periodic attractor $\mathcal{P}_s$ (cyan surface), and the homoclinic bifurcation curve (light green curve) have the same meaning as in Fig. \ref{fig:singular-orbit}. The superslow manifold $M_{ss}$ changes stability from attracting (red solid) to repelling (red dashed) at the HB (red diamond). Along the lower fold curve, a folded node (FN, blue circle) is located $O(\delta)$ close to the CDH (yellow square). The magenta circle is a full-system equilibrium (TE), which is a saddle-focus. (B) Two-dimensional (2D) cross-sections of the $M_s$ and $\mathcal{P}_s$ in the $(n,V)$-space, with $q$ fixed at its minimum (cyan curves) and maximum (orange curves) values. (C) Projection of the curves of the fold $L_s$, $M_{ss}$, and the HC, together with FN, CDH, and HB, TE points, and the trajectory onto the $(n,q)$-space.}
\label{fig:nqv-default}
\end{figure}

At $r_\varepsilon=r_\delta=1$, the full system \eqref{eq:slow-theta} generates MMOs consisting of one large-amplitude oscillation (LAO) and four small-amplitude oscillations (SAOs) per cycle (Fig. \ref{fig:bd}D, top row). The geometric configuration of this MMO trajectory projected onto $(n,q,V)$-space is shown in Fig.~\ref{fig:nqv-default}A. The critical manifold $M_s$ (blue surface), fold curves $L_s$ (blue curves), the superslow manifold $M_{ss}$ (red curve), the homoclinic bifurcation curve (HC, light green curve), and the periodic attractor $\mathcal{P}_s$ (cyan surface) have the same meanings as in Fig.~\ref{fig:singular-orbit}. The $M_{ss}$ consists of an attracting branch $M_{ss}^a$ (red solid) and a repelling branch $M_{ss}^a$ (red dashed), which are separated by the DHB (red diamond). The folded singularity is a folded node (FN, blue circle), which occurs $O(\delta)$ close to the CDH (yellow square). The full-system equilibrium (TE, magenta circle) lying on the middle sheet of $M_s$ is a saddle-focus. Fig. \ref{fig:nqv-default}B shows 2D cross-sections of the 3D surfaces $M_s$ and $\mathcal{P}_s$ in $(n,V)$-space, obtained by fixing $q$ at its minimum (cyan curves) and maximum (orange curves).  Fig. \ref{fig:nqv-default}C shows the projection of the bifurcation curves, key singularity points, and the MMO trajectory from panel (A) onto $(n,q)$-space. 

Starting near the attracting side of $M_{ss}$ at the maximum value of $q$, the trajectory oscillates around $M_{ss}^a$ while traveling toward the DHB as $q$ decreases, during which the SAOs gradually decrease in magnitude. Upon crossing the DHB onto $M_{ss}^r$, the trajectory spirals out with growing amplitudes before it jumps away to fire an action potential, which is then terminated upon crossing the HC bifurcation. Following the spike, the trajectory reverses direction and travels toward increasing $q$, returning to the maximal value of $q$ and completing a full MMO cycle. Thus, in the default case, the MMOs are organized by the DHB mechanism predicted by the GSPT analysis. Although the FN, CDH, and TE points are all present, they remain relatively far away from the SAO region and therefore do not play a significant role. 

\subsection{MMOs transition to MMBOs via exaggerating timescale separation}\label{sub:exaggeration-theta}
\begin{figure}[!htp]
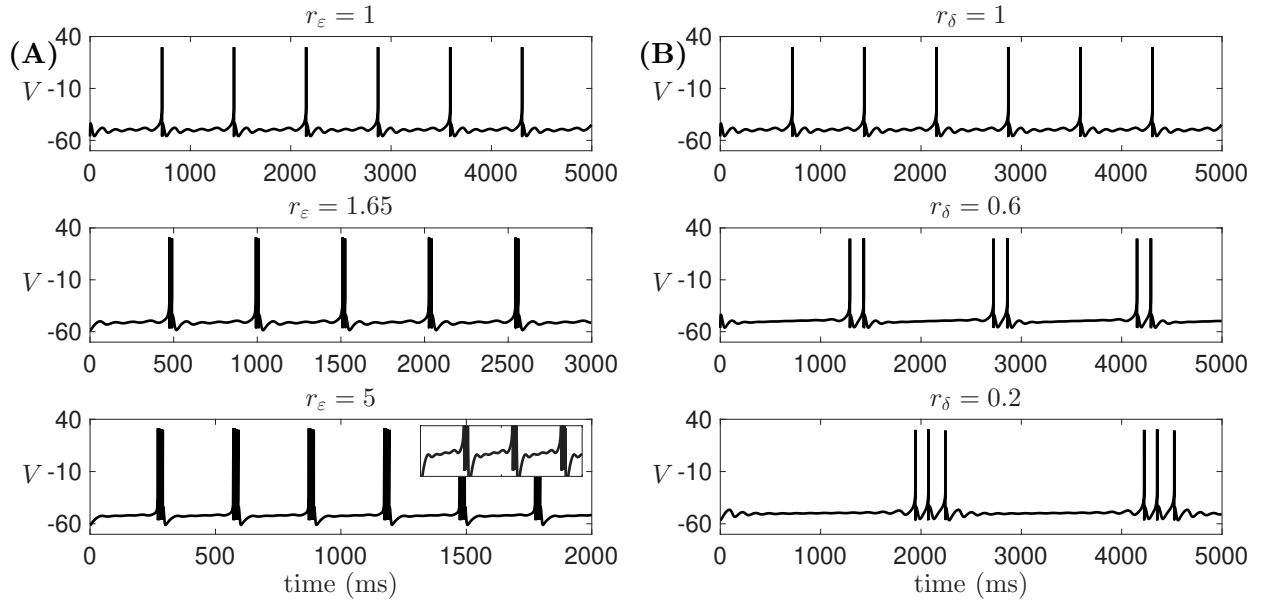

\begin{center}
\begin{tabular}
{@{}p{0.48\linewidth}@{\quad}p{0.48\linewidth}@{}}
\subfigimg[width=\linewidth]{\makebox[0pt][l]{\raisebox{1.2ex}[0pt][0pt]{\bfseries{\small{(A)}}}}}{fix_delta_increasing_eps.eps} &
\subfigimg[width=\linewidth]{\makebox[0pt][l]{\raisebox{1.2ex}[0pt][0pt]{\bfseries{\small{(B)}}}}}{fix_eps_decreasing_delta.eps}
\end{tabular}
\end{center}
\caption{Time traces of \eqref{eq:slow-theta} illustrating MMO-MMBO transitions as the three-timescale separation increases. (A) For fixed $r_\delta = 1$, increasing $r_\varepsilon$ from  $r_\varepsilon = 1$ (top) to $r_\varepsilon = 1.65$ (middle) and $r_\varepsilon = 5$ (bottom). (B) For fixed $r_\varepsilon = 1$, decreasing $r_\delta$ from $r_\delta = 1$ (top) to $r_\delta = 0.6$ (middle) and $r_\delta = 0.2$ (bottom). Other parameters are given in Tables \ref{tab:para-theta} and \ref{tab:activation}.}
\label{fig:exaggeration-timetraces}
\end{figure}

We next consider the transitions from the default MMOs to MMBOs that occur when the system is driven closer to its double singular limit.
As illustrated in Fig.~\ref{fig:all-transitions}A, this can be achieved by decreasing either $\varepsilon$ or $\delta$ from the default parameter values (magenta star). See Fig.~\ref{fig:exaggeration-timetraces} for representative solutions. 

We analyze the two transitions separately and compare the resulting MMBO dynamics with the theoretical predictions obtained from the singular-limit analysis. When $\delta$ is fixed at the default value and $\varepsilon$ is decreased (Fig.~\ref{fig:exaggeration-timetraces}A), the transition from MMOs to MMBOs can occur along the same isola so that the number of SAOs is preserved, but the SAO-generating mechanism changes from DHB-dominated to CDH. This route leads to the perturbed trajectory in Fig.~\ref{fig:singular-orbit}D when $\varepsilon$ is reduced to $O(0.01)$. In contrast, when $\varepsilon$ is fixed and $\delta$ is decreased (Fig.~\ref{fig:exaggeration-timetraces}B), the transition occurs across different isolas, while the SAOs remain DHB-dominated. This second route leads to MMBOs that are analogous to the perturbed trajectory in Fig.~\ref{fig:singular-orbit}C.  

\subsubsection{Fixing $r_\delta=1$ and speeding up fast variables}

\begin{figure}[!htp]
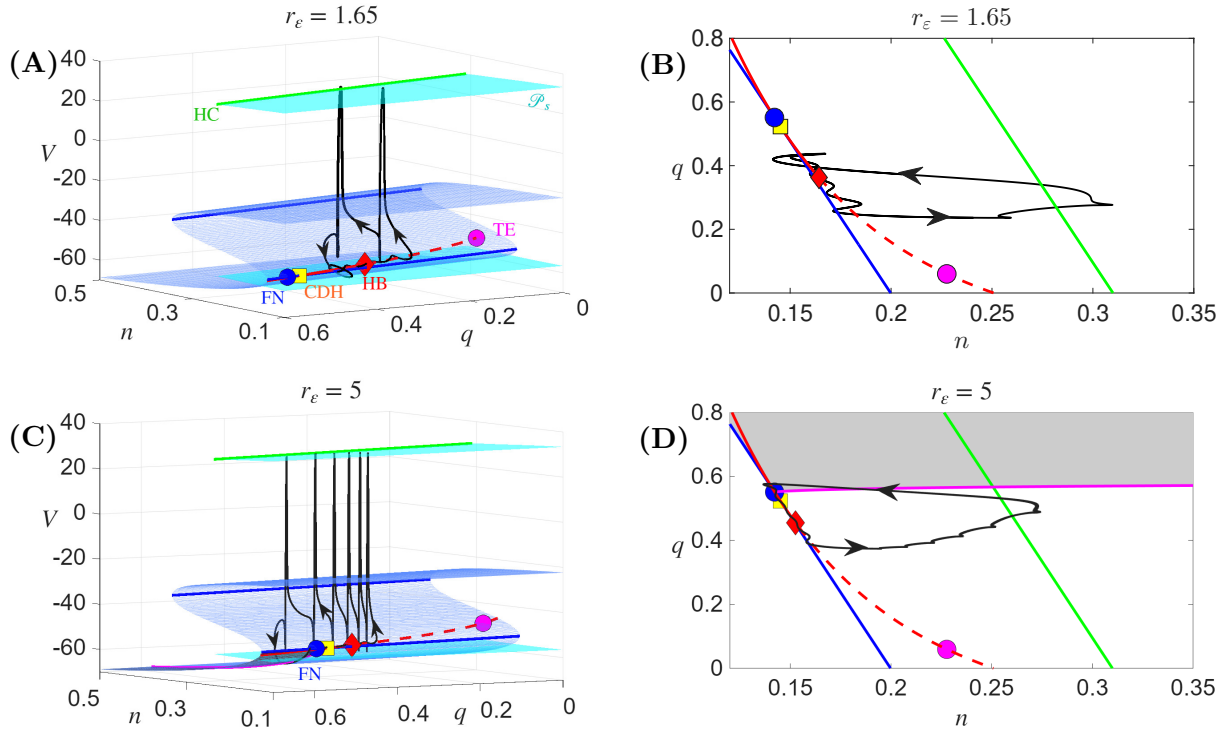

\begin{center}
\begin{tabular}
{@{}p{0.48\linewidth}@{\quad}p{0.48\linewidth}@{}}
\subfigimg[width=\linewidth]{\bfseries{\small{(A)}}}{projection_nqv_eps_1p65.eps} &
\subfigimg[width=\linewidth]{\bfseries{\small{(B)}}}{twopar_hc_nq_eps_1p65.eps}\\
\subfigimg[width=\linewidth]{\bfseries{\small{(C)}}}{projection_nqv_i8_r5.eps} &
\subfigimg[width=\linewidth]{\bfseries{\small{(D)}}}{twopar_hc_nq_r5.eps}
\end{tabular}
\end{center}
\caption{Projections of the periodic MMBO solutions of \eqref{eq:slow-theta} from Fig.~\ref{fig:exaggeration-timetraces}A onto $(n,q,V)$-space (left panels) and $(n,q)$-plane (right panels), together with the corresponding geometric structures, for $r_\delta = 1$ and 
(A, B) $r_\varepsilon=1.65$ 
and (C, D) $r_\varepsilon=5$. 
Increasing $r_\varepsilon$ shifts the SAO mechanism from the DHB (A, B) to the CDH (C, D) by bringing the DHB (red diamond) closer to the CDH (yellow square), allowing the trajectory to enter the funnel (shaded region) of the FN (blue circle) before passing through the CDH and the DHB. Other colors and symbols have the same meaning as in Figs. \ref{fig:singular-orbit} and \ref{fig:nqv-default}.} 
\label{fig:nqv-increasing-eps}
\end{figure} 

As illustrated in Section \ref{sec:numerical-simu-theta}, increasing $r_\varepsilon$ (equivalently, decreasing $\varepsilon$) along the same isola leads to a transition from MMOs to MMBOs via a spiking-adding cascade, during which the number of spikes (i.e., LAOs) gradually increases yet the number of SAOs remains unchanged. Without loss of generality, we focus on the isola formed by 4 SAOs. Representative solutions along this branch are shown in Fig.~\ref{fig:exaggeration-timetraces}A and projected onto $(n,q,V)$- and $(n,q)$-spaces in Fig.~\ref{fig:nqv-increasing-eps}. 

As $r_{\varepsilon}$ increases, the fast variables evolve more rapidly relative to the slow variables, thereby allowing the system to generate more spikes during the active phase. At the same time, decreasing $\varepsilon$ brings the system closer to the $\varepsilon\to 0$ singular limit, so the critical manifold $M_s$, the folded node (FN), and the CDH singularity become increasingly important in organizing the subthreshold dynamics. Moreover, because the DHB point is $O(\varepsilon)$ close to the CDH, it moves closer to the CDH as $\varepsilon$ decreases. As a result, the mechanism underlying the SAOs shifts from a pure DHB mechanism to a CDH mechanism involving an interaction between the DHB and the canard dynamics of the folded node. This transition is illustrated in Fig.~\ref{fig:nqv-increasing-eps}: for $r_\varepsilon=1.65$, the MMBO remains DHB-mediated, similar to the default case analyzed above, whereas for larger $r_\varepsilon$ values, the MMBO becomes CDH-mediated (see Figs. \ref{fig:nqv-increasing-eps}C, D for $r_\varepsilon=5$ and Fig.~\ref{fig:singular-orbit}D for $r_\varepsilon=10.397$). As illustrated previously in Section \ref{sub:singular-orbit}, a CDH-mediated MMBO trajectory enters the funnel region of the folded node (Fig. \ref{fig:nqv-increasing-eps}D, gray shaded region) and is guided toward the folded node before passing through the CDH and DHB onto the repelling branch.


Thus, along the same isola branch where the number of SAOs is preserved, decreasing $\varepsilon$ changes the SAO-generating mechanism. More generally, decreasing $\varepsilon$ can also move the dynamics to different isola branches and thereby change the number of SAOs; nevertheless, the SAO mechanism follows the same trend. As $\varepsilon$ decreases with $\delta$ fixed at its default value, the SAOs shift from a DHB-dominated mechanism to a CDH mechanism involving both DHB and folded-node canard dynamics.



\subsubsection{Fixing $r_\varepsilon=1$ and slowing down the superslow variable}


\begin{figure}[!t]
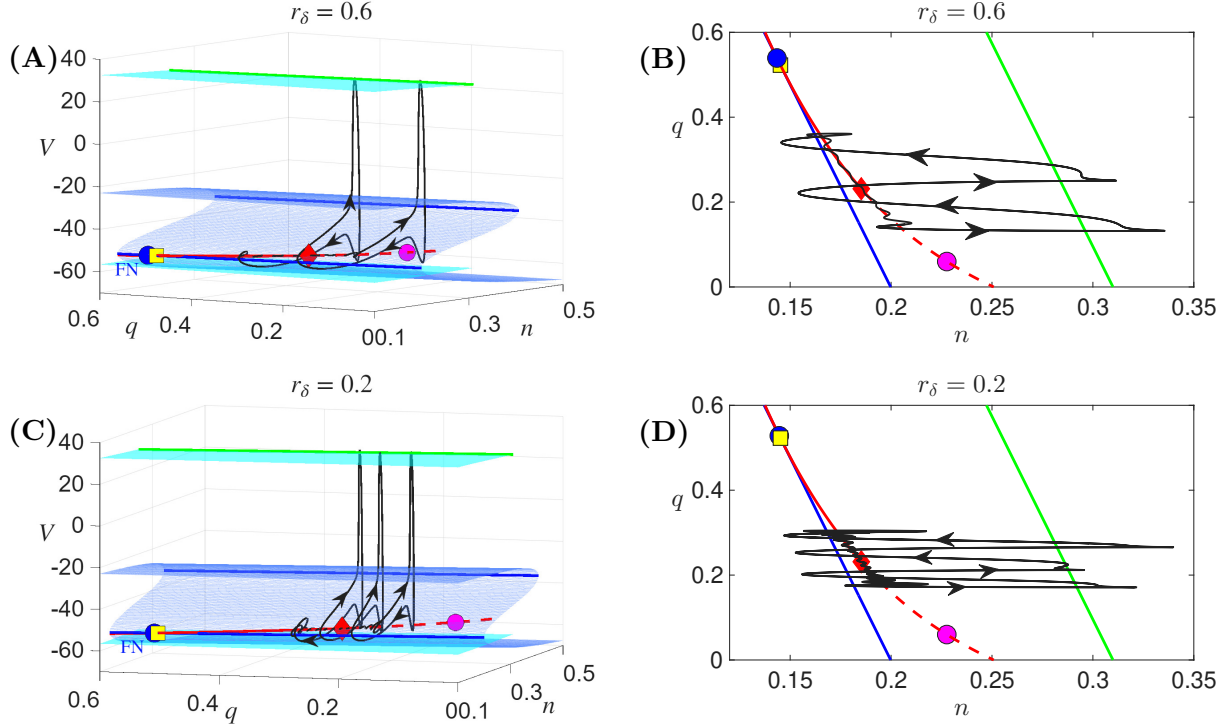

\begin{center}
\begin{tabular}
{@{}p{0.48\linewidth}@{\quad}p{0.48\linewidth}@{}}
\subfigimg[width=\linewidth]{\bfseries{\small{(A)}}}{projection_nqv_p0p6.eps} &
\subfigimg[width=\linewidth]{\bfseries{\small{(B)}}}{twopar_hc_nq_p06.eps}\\
\subfigimg[width=\linewidth]{\bfseries{\small{(C)}}}{projection_nqv_0p2.eps} &
\subfigimg[width=\linewidth]{\bfseries{\small{(D)}}}{twopar_hc_nq_p02.eps}
\end{tabular}
\end{center}
\caption{Projections of the attracting MMBO solutions of \eqref{eq:slow-theta} from Fig. \ref{fig:exaggeration-timetraces}B onto $(n,q,V)$-space (left panels) and $(n,q)$-plane (right panels), together with the corresponding geometric structures, for $r_\varepsilon= 1$ and (A, B) $r_\delta=0.6$ 
and (C, D) $r_\delta=0.2$. 
Decreasing $r_\delta$ leads to increasingly pronounced DHB-like features. Although the FN (blue circle) moves closer to the CDH (yellow square), the trajectory remains outside the funnel of the FN and the neighborhood of the CDH. Other colors and symbols have the same meaning as in Figs. \ref{fig:singular-orbit} and \ref{fig:nqv-default}.} 
\label{fig:nqv-decreasing-delta}
\end{figure}

We now consider the transition from MMOs to MMBOs induced by decreasing $r_\delta$ (equivalently, decreasing $\delta$), which slows the evolution of the superslow variable. Unlike the $\varepsilon$-induced MMBOs, for which multiple spikes occur within a single burst episode, the MMBO induced by decreasing $\delta$ consists of multiple burst episodes, each containing only a single spike that initiates at a regular fold point of $M_s$ and terminates after crossing the HC curve (see Fig.~\ref{fig:nqv-decreasing-delta}). 
 
For smaller values of $\delta$, the slower evolution of the superslow variable $q$ allows the trajectory to spend more time evolving along $M_{ss}$ during the SAO phase. This leads to a more pronounced delay after passage through the DHB and more prominent DHB-like features in the SAOs. The same slowing of $q$ also promotes the emergence of MMBOs: during the active phase, the slower increase of $q$ enables the trajectory to undergo repeated jumps at regular fold points, thereby generating multiple burst episodes before returning to the SAO phase. This dynamics is analogous to the MMBO shown in Fig.~\ref{fig:singular-orbit}C, except that each multi-spike burst is replaced here by a single-spike burst. This difference arises because, for the default values of $\varepsilon$ used here (Figs.~\ref{fig:nqv-default} and~\ref{fig:nqv-decreasing-delta}), each burst episode contains only a single spike before termination by the HC curve. By contrast, the MMBO in Fig.~\ref{fig:singular-orbit}C is closer to the singular limit in both $\varepsilon$ and $\delta$ values: the smaller value of $\varepsilon$ allows multiple spikes to occur within each burst episode, while the smaller value of $\delta$ allows more burst episodes to occur before the trajectory returns to the SAO phase.

Although decreasing $r_\delta$ brings the FN closer to the CDH, both points remain far away from the trajectory and hence do not contribute to organizing the SAOs. Thus, the SAO mechanism remains DHB-dominated. This observation is consistent with the singular-limit predictions discussed in Section \ref{sub:singular-orbit}.


\subsection{MMOs are robust to speeding up the superslow variable}\label{sub:mmos-preservation-theta} 

Having understood the effect of parameter changes that exaggerate timescale separation, we now consider the opposite direction: increasing either $\varepsilon$ or $\delta$ pushes the system away from the singular limit and hence weakens the relevance of the geometric structures identified in Section \ref{sec:gspt-theta}. 
The full system bifurcation diagram in Fig.~\ref{fig:all-transitions} shows that MMOs nevertheless persist over a substantial region of parameter space. 

For increasing $\varepsilon$, MMOs persist until giving way to purely subthreshold oscillations after a torus bifurcation. As $\varepsilon$ increases further, the system eventually reaches a quiescent state after a Hopf bifurcation. For increasing $\delta$, the robustness of MMO dynamics is stronger: MMOs remain robust even after the system enters a regime where canard and DHB mechanisms are no longer dynamically relevant. We show below that this occurs because increasing $\delta$ brings the trajectory closer to the full-system saddle-focus equilibrium TE, which can organize local SAOs \cite{Desroches2012, Kugler2016,Phan2023}. 



\begin{figure}[!htp]
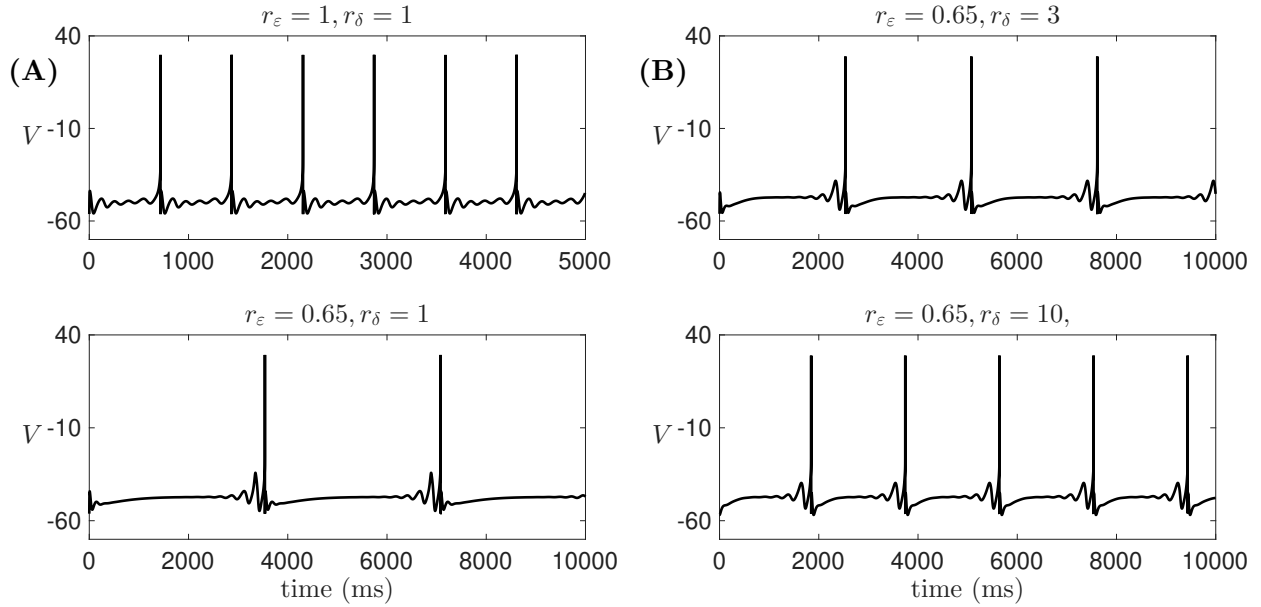

\begin{center}
\begin{tabular}
{@{}p{0.48\linewidth}@{\quad}p{0.48\linewidth}@{}}
\subfigimg[width=\linewidth]{\bfseries{\small{(A)}}}{fix_delta_decreasing_eps.eps} &
\subfigimg[width=\linewidth]{\bfseries{\small{(B)}}}{fix_eps_p65_increasing_delta.eps}
\end{tabular}
\end{center}
\caption{Time traces of \eqref{eq:slow-theta} illustrating the persistence of MMOs as the three-timescale separation is weakened. (A) For fixed $r_\delta = 1$, decreasing $r_\varepsilon$ from $r_\varepsilon = 1$ (top) to $r_\varepsilon = 0.65$ (bottom). (B) For fixed $r_\varepsilon = 0.65$, increasing $r_\delta$ from $r_\delta = 3$ (top) to $r_\delta = 10$ (bottom).
} 
\label{fig:mmos-robust}
\end{figure}

\subsubsection{Fixing $r_\delta=1$ and slowing down the fast variables}\label{sub:delta-1-decreasing-eps}

We first move away from the double singular limit by decreasing $r_\varepsilon$ with $r_\delta$ fixed (Fig. \ref{fig:mmos-robust}A). This slows down the fast variables, thereby bringing the system closer to a 7-slow/1-superslow (7S, 1SS) decomposition. Thus, the critical manifold $M_s$ and folded node becomes less meaningful, while the superslow manifold $M_{ss}$ and the DHB become more important for organizing the dynamics. 

\begin{figure}[!htp]
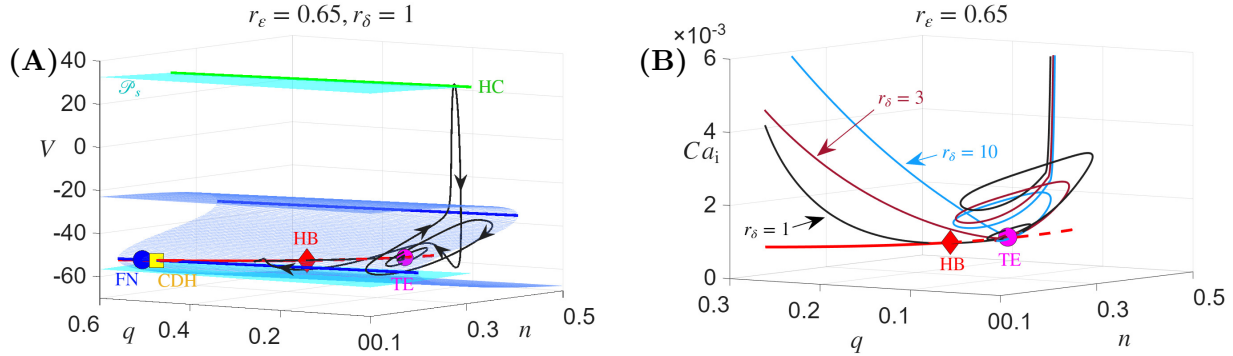

\begin{center}
\begin{tabular}
{@{}p{0.48\linewidth}@{\quad}p{0.48\linewidth}@{}}
\subfigimg[width=\linewidth]{\bfseries{\small{(A)}}}{nqv_r_0p65.eps} &
\subfigimg[width=\linewidth]{\bfseries{\small{(B)}}}{n_q_ca_projection_eps_0p65.eps}
\end{tabular}
\end{center}
\caption{Projections of the attracting MMO solutions of \eqref{eq:slow-theta} from Fig. \ref{fig:mmos-robust} onto (A) $(n,q,V)$-space and (B) $(n,q,Ca_{\mathrm{i}})$-space, together with the corresponding geometric structures, for fixed $r_\varepsilon = 0.65$ and different values of $r_\delta$. The black curve represents the projection of the trajectory from Fig. \ref{fig:mmos-robust}A (bottom) with $r_\delta = 1$. The brown and cyan curves indicate the projections of the trajectories from Fig. \ref{fig:mmos-robust}B with $r_\delta = 3$ and $r_\delta = 10$, respectively. Other colors and symbols have the same meaning as in Figs. \ref{fig:singular-orbit} and \ref{fig:nqv-default}.
} 
\label{fig:nqv-eps-0p65}
\end{figure}

As $r_\varepsilon$ decreases (i.e., $\varepsilon$ increases), the DHB shifts away from the CDH and therefore closer toward the TE (compare Fig. \ref{fig:nqv-default}A and Fig.~\ref{fig:nqv-eps-0p65}A). As a result, the DHB-induced delay along $M_{ss}^r$ is now able to bring the orbit into the neighborhood of the saddle-focus TE (see the black trajectory in Fig. \ref{fig:nqv-eps-0p65}). Once in this neighborhood, the trajectory follows the two-dimensional unstable manifold of the TE (not shown), producing SAOs with growing amplitude as it spirals away. Thus, in this regime, the SAO arises from the combined influence of the DHB and the saddle-focus TE. 


\subsubsection{Fixing $r_\varepsilon=0.65$ and speeding up the superslow variable}

Having identified the mechanism underlying the MMO in Fig.~\ref{fig:nqv-eps-0p65}A when $r_\varepsilon = 0.65$ and $r_\delta=1$, we now proceed to examine the robustness of these MMOs to increasing $r_\delta$. Increasing $r_\delta$ (equivalently, increasing $\delta$) speeds up the superslow variable and, for sufficiently large values such as $r_\delta = 10$, eliminates the effective timescale separation. 
As a result, the critical manifold, superslow manifold, folded singularity, DHB, and CDH no longer organize the dynamics. Nonetheless, MMOs persist, suggesting that the SAOs are organized instead by the full-system equilibrium TE, which remains a saddle-focus as $r_\delta$ increases.

Fig. \ref{fig:mmos-robust}B shows the effect of increasing $r_\delta$ on time traces, with the corresponding projections onto $(n,q, Ca_{\mathrm{i}})$-space shown in Fig. \ref{fig:nqv-eps-0p65}B. 
As $r_\delta$ increases, the trajectories are gradually pulled away from the $M_{ss}$ and no longer pass through the DHB (Fig.~\ref{fig:nqv-eps-0p65}B, brown and cyan trajectories). Instead, they directly approach the saddle-focus TE along its stable manifold and then generate SAOs with increasing amplitude as they spiral away along its unstable manifold. Consequently, in a regime away from the singular limit, the SAOs in these MMOs are induced solely by the saddle-focus TE mechanism. 

\begin{remark}
This same saddle-focus mechanism that underlies the robustness of MMOs to increasing $r_{\delta}$ (Fig.~\ref{fig:all-transitions}A, cyan region) also explains the persistence of MMBO dynamics over a wide range of $r_\delta$ values for intermediate $r_\varepsilon$ values (Fig.~\ref{fig:all-transitions}A, upper-left yellow region). We therefore omit a separate analysis of these MMBOs. 
\end{remark}

\subsection{From MMBOs to regular bursting via speeding up fast variables or superslow variable}\label{sub:mmbos-bursting}

The final transitions considered in this subsection explain how MMBOs lose their SAOs and become regular bursting solutions. We consider two representative MMBO regimes from Fig.~\ref{fig:all-transitions}A. The first lies in the upper-left yellow region, where SAOs are organized by the saddle-focus TE. In this case, the transition to bursting is induced by increasing $r_\varepsilon$. The second lies in the lower-right yellow region, where SAOs are organized by a three-timescale CDH mechanism and the transition to bursting is induced by increasing $r_\delta$. 

\paragraph{Fixing $r_\delta=10$ and increasing $r_\varepsilon$}
In the upper-left yellow region of Fig.~\ref{fig:all-transitions}A, the MMBOs persist for large $\varepsilon$ and $\delta$ values even though the singular structures no longer organize the SAOs. 
As explained above, these SAOs are instead generated by the full-system saddle-focus equilibrium TE. As $r_\varepsilon$ increases from this regime, the MMBO quickly transitions to regular bursting because the TE changes from a saddle-focus equilibrium to a saddle equilibrium with real eigenvalues of both signs. Note that increasing $r_\varepsilon$ speeds up the fast variables, thereby moving the system closer to a 6-fast/2-slow (6F, 2S) decomposition, which could potentially restore the influence of folded singularities. Nonetheless, at relatively large values of $r_\delta$, the folded singularity is a folded focus (Fig.~\ref{fig:twopar-hopf-theta}A) and therefore does not support canard dynamics. Thus, this transition is induced by the loss of the full-system saddle-focus mechanism: once TE no longer supports local spiraling, and with neither folded-node canards nor DHB playing a role at large values of $r_\delta$, the system exhibits regular bursting dynamics without SAOs.


\paragraph{Fixing $r_\varepsilon=7.5$ and increasing $r_\delta$}

\begin{figure}[!htp]
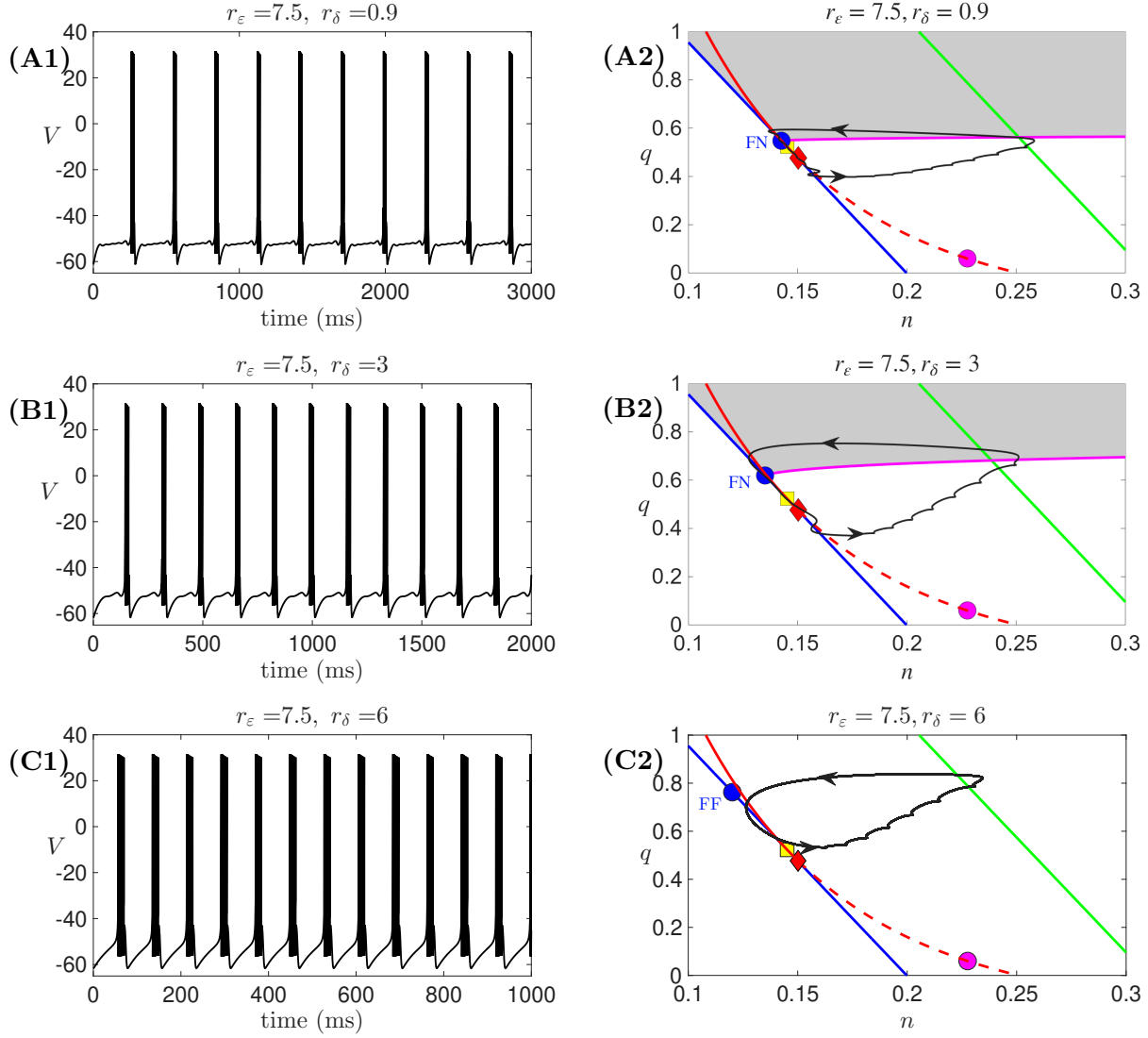

\begin{center}
\begin{tabular}
{@{}p{0.48\linewidth}@{\quad}p{0.48\linewidth}@{}}
\subfigimg[width=\linewidth]{\bfseries{\small{(A1)}}}
{timetrace_eps_7p5_del_0p9.eps} &
\subfigimg[width=\linewidth]{\bfseries{\small{(A2)}}}{nq_eps_7p5_delta_0p9.eps}\\
\subfigimg[width=\linewidth]{\bfseries{\small{(B1)}}}
{timetrace_eps_7p5_del_3.eps} &
\subfigimg[width=\linewidth]{\bfseries{\small{(B2)}}}{nq_eps_7p5_delta_3.eps}\\
\subfigimg[width=\linewidth]{\bfseries{\small{(C1)}}}
{timetrace_eps_7p5_del_6.eps} &
\subfigimg[width=\linewidth]{\bfseries{\small{(C2)}}}{nq_eps_7p5_delta_6.eps}
\end{tabular}
\end{center}
\caption{Time traces of solutions of \ref{eq:slow-theta} (left panels) and their projections onto $(n, q)$-plane together with the corresponding geometric structures (right panels) for $r_\varepsilon = 7.5$, (A) $r_\delta = 0.9$, (B) $r_\delta = 3$, and (C) $r_\delta = 6$. Increasing $r_\delta$ produces SAOs with more pronounced canard-like features while gradually weakening DHB characteristics. When $r_\delta$ is sufficiently large (e.g., $r_\delta = 6$), the folded node FN transitions to a folded focus FF, leading to the loss of SAOs. Other colors and symbols have the same meaning as in Fig. \ref{fig:nqv-default}C. 
} 
\label{fig:nqv-mmbos-bursting}
\end{figure}

The second MMBO-to-bursting transition begins in the lower-right region of Fig.~\ref{fig:all-transitions}A, where the MMBOs are organized by genuine three-timescale CDH. We examine this transition for fixed $r_\varepsilon = 7.5$ and increasing $r_\delta$ (Fig. \ref{fig:nqv-mmbos-bursting}). 

At $r_\varepsilon = 7.5$ and $r_\delta=0.9$, the system lies in a pronounced three-timescale setting, where MMBOs are CDH-induced, with both the folded node and DHB organizing the SAO dynamics (Fig. \ref{fig:nqv-mmbos-bursting}A). When $r_\delta$ is increased to $3$ (Fig. \ref{fig:nqv-mmbos-bursting}B), the dynamics remains CDH-governed, but the folded node moves farther away from the CDH and the DHB contribution weakens, as reflected by the reduced symmetry of the SAOs with respect to the DHB. 
Further increasing $r_\delta$ to $6$ (Fig. \ref{fig:nqv-mmbos-bursting}C) moves the folded singularity even farther away from both the CDH and the trajectory. Moreover, the folded node becomes a folded focus (FF), while the superslow manifold and DHB no longer organize the dynamics. As a result, both the DHB and folded-node canard mechanisms are lost. Since the TE point remains far away from the trajectory, the system generates bursting dynamics without SAOs.

\section{Discussion}\label{sec:discussion-theta}

In this article, we combined geometric singular perturbation theory (GSPT) with full-system bifurcation analysis to study mixed-mode bursting dynamics (MMBOs) in an eight-dimensional biophysical model of cortical theta neuronal oscillators with three distinct timescales. Previous geometric analyses of MMBOs have largely focused on two timescale reductions, in which delayed-Hopf bifurcation (DHB) and folded-node canard mechanisms are treated as separate mechanisms. The three-timescale formulation studied here provides a unified setting in which these mechanisms can coexist, interact and exchange dominance near a three-timescale canard-delayed-Hopf (CDH) singularity as the relative timescale separations are varied. Full-system bifurcation analysis further revealed how MMBOs, MMOs, and regular bursting are organized over a much larger range of $\varepsilon,\delta>0$, including regimes beyond the singular-limit predictions, where a full-system saddle-focus equilibrium provides an additional SAO-generating mechanism. Together, these analyses show that MMBOs in this system are not organized by a single mechanism, but by a combination of DHB, folded-node canard, CDH, and saddle-focus effects whose relative roles depend on the timescale structure. 

The GSPT analysis provides a geometric framework for interpreting MMBO dynamics when $0<\varepsilon,\delta\ll 1$. In this regime, the two timescale parameters play complementary roles. The parameter $\varepsilon$ determines the fast-slow timescale separation, the relative importance of the folded-node canard mechanism, and the distance between the DHB and the CDH, whereas $\delta$ controls the drift of the superslow variable, the relative importance of the DHB mechanism, and the distance between the folded singularity and the CDH. In the double singular limit, the model admits a continuum of singular bursting orbits over the interval $0\le q \le q_{\mathrm{CDH}}$ (Fig.~\ref{fig:singular-orbit}A). For fixed small $\varepsilon$, the singular orbits perturb to distinct MMBO regimes as $\delta$ is increased. For very small $\delta$, the perturbed MMBO trajectory remain away from the folded node and CDH, and the SAOs are primarily DHB-mediated (Fig.~\ref{fig:singular-orbit}C). For larger but still small $\delta$, the trajectory enters the funnel region of the folded node and passes through the CDH, producing MMBOs with both DHB and folded-node canard characteristics (Fig.~\ref{fig:singular-orbit}D). Thus, within the GSPT regime, increasing $\delta$ shifts MMBOs from a DHB-dominated regime to a CDH-mediated regime by bringing the folded-node canard mechanism into play, whereas increasing $\varepsilon$ weakens and separates the folded-node canard mechanism from the DHB, thereby making the DHB mechanism more dominant. 

The full-system bifurcation analysis complements the GSPT by showing how MMBOs are organized over broader ranges of $\varepsilon,\delta>0$. Continuation with respect to $\varepsilon$ shows that periodic MMBO solutions are organized along families of isolas, revealing a sequence of spike-adding SNPO and PD bifurcations (Fig.~\ref{fig:bd}). Isolas of MMBOs were previously computed in a two slow/two fast model \cite{koksal2020canard}, where each isola contained a single stable $L$-spike MMBO segment. In contrast, the isolas in the present three-timescale conductance-based model contain multiple stable segments bounded by PD and SNPO bifurcations. These segments correspond to $L$-spike MMBOs with different values of $L$, while the number of SAOs remains fixed along each isola. Thus, as $\varepsilon$ decreases, an isola associated with a given number of SAOs supports a spike-adding cascade through which stable $L$-spike MMBOs give way to stable $(L+1)$-spike MMBOs.

Two-parameter full-system bifurcation analysis further reveals how the structure of these isolas depends on the superslow timescale parameter $\delta$. As $\delta$ decreases, individual isolas extend to smaller values of $\varepsilon$ and develop a more complicated structure that contains more stable MMBO segments. As $\delta$ increases, the isola structure progressively simplifies: stable MMBO segments with more spikes are eliminated first, leaving fewer stable segments along each isola. 
Eventually, the remaining isola supports only a single stable segment of $1^s$ orbits bounded by two PD bifurcations, before the entire isola is destroyed through a cusp bifurcation of PD bifurcations (Fig.~\ref{fig:all-transitions}B). Consequently, for small $\delta$, transitions between MMOs and MMBOs can be induced by different bifurcations because many isolas overlap in parameter space. For larger $\delta$, the structure becomes simpler, and the SNPO curve associated with the $1^1$ MMO branch provides a close approximation to the MMO/MMBO transition boundary (Fig.~\ref{fig:all-transitions}A).

By combining the GSPT and full-system bifurcation analysis, we further clarified how changes in timescale separation drive transitions among MMOs, MMBOs, and bursting dynamics. At the default timescale parameters, the model exhibits MMOs that are primarily DHB-mediated, with the CDH and folded node located relatively far away from the trajectory (Fig.~\ref{fig:nqv-default}). From this reference point in the $(\varepsilon,\delta)$ parameter space, we showed that exaggerating the three-timescale separation (i.e., pushing the system closer to the double singular limit) can induce the transition from MMOs to MMBOs. This can be achieved by either speeding up the fast variables through decreasing $\varepsilon$ or slowing down the superslow variable via decreasing $\delta$. These two routes both increase the effective separation of timescales and allow additional spikes to occur, but they affect the SAO mechanism differently. Decreasing $\varepsilon$ brings the DHB point closer to the CDH and increases the role of folded-node canard dynamics, thereby inducing a transition in the SAO-generating mechanism from primarily DHB-driven to CDH-mediated dynamics  (Fig.~\ref{fig:nqv-increasing-eps}). Because this transition can occur along a single isola, the number of SAOs can remain fixed even as the underlying SAO mechanism changes. 
In contrast, decreasing $\delta$ enhances the DHB mechanism and keeps the trajectory farther away from the CDH and folded-node, so that the MMBOs remain DHB-driven (Fig.~\ref{fig:nqv-decreasing-delta}).

Moving the system away from the singular limit reveals additional mechanisms that are not captured by the GSPT analysis alone. Increasing $\varepsilon$ from the default regime moves the system toward an effective seven slow/one superslow configuration and promotes the importance of the DHB mechanism.  At the same time, it brings the trajectory closer to the full-system saddle-focus equilibrium (TE), so that SAOs can be organized by a combination of DHB and saddle-focus effects (Fig.~\ref{fig:nqv-eps-0p65}A). For larger $\delta$, the DHB becomes dynamically less relevant, and the SAOs are governed primarily by the TE mechanism (Fig.~\ref{fig:nqv-eps-0p65}B). Because the location of the TE is independent of $\varepsilon$ and $\delta$, TE-mediated MMOs and MMBOs can be remarkably robust to increasing $\delta$. 

The transition from MMBOs to regular bursting without SAOs occurs through two distinct pathways. In regimes where the SAOs are organized by the full-system saddle-focus equilibrium (Fig.~\ref{fig:all-transitions}A, upper-left yellow region), decreasing $\varepsilon$ changes the eigenvalue configuration of the TE from a saddle-focus to a saddle with real eigenvalues. Once the saddle-focus effects are lost, the trajectory no longer produces SAOs and regular bursting emerges. In contrast, in the genuine three-timescale CDH regime with small $\varepsilon$ and $\delta$ values (Fig.~\ref{fig:all-transitions}A, lower-right yellow region), increasing $\delta$ weakens the dynamical role of the DHB and moves the folded singularity away from the CDH. Eventually, the folded node becomes a folded focus for $\delta$ large enough, eliminating the canard mechanism, while the DHB no longer organize the local dynamics (Fig.~\ref{fig:nqv-mmbos-bursting}). Moreover, in this regime, TE remains far from the dynamics and is a saddle equilibrium, as explained above. Thus, for small $\varepsilon$ and large $\delta$, the system exhibits regular bursting with no SAOs.

We have computed, for the first time, a detailed bifurcation diagram of MMBOs in a three-timescale model and identified a richer isola structure than previously reported in two-timescale MMBO models \cite{koksal2020canard}. In their system, spike-adding along an MMBO isola occurred through canard-mediated transitions that affected both the number of SAOs and the number of large-amplitude spikes per burst, with the SAOs arising from a single folded-node canard mechanism. In our system, by contrast, the underlying SAO-generating mechanism can vary along MMBO isolas. A detailed analysis of the solution profiles and spike-adding mechanisms along these isolas remains to be carried out.  Beyond the intrinsic dynamics studied here, it would also be interesting to determine whether the dynamical mechanisms identified in this work have functional consequences for the ability of the oscillator to phase-lock to external forcing. The default MMO regime of the theta oscillator \eqref{eq:main-theta} is governed by a DHB mechanism, which has been shown to generate prolonged post-input recovery delays that greatly expand the range of input frequencies over which entrainment can occur \cite{Pittman2021,wang2026dynamical}. Whether the additional mechanisms identified here, including CDH and folded-node canards, similarly contribute to phase-locking flexibility is left to future work.

\section*{Acknowledgments}

The authors wrote the original manuscript. Portions of original manuscript text were submitted to an AI large language model to improve readability and grammar. After
review and revision, some of these passages were included.

\begin{appendix}

\section{Parameters of the cortical theta oscillator model \eqref{eq:main-theta}}\label{sec:para-theta}

This section lists the parameter values used in model~\eqref{eq:main-theta}; see Tables~\ref{tab:para-theta} and~\ref{tab:activation}.

\begin{table}[!t]
\caption{Parameters in the model \eqref{eq:main-theta}}
\centering
\begin{tabular}{|c c c|c c c|}
\hline
$g_{\mathrm {Na}}$  & $125$ & $\mathrm{mS/cm^2}$ & $E_{\mathrm {Na}}$ & $40$ & $\mathrm{mV}$ \\

$g_{\mathrm {K_{DR}}}$ & $54$ & $\mathrm{mS/cm^2}$ & $E_{\mathrm K}$ & $-80$ & $\mathrm{mV}$\\

$g_{\mathrm {leak}}$ & $0.27$ & $\mathrm{mS/cm^2}$ & $E_{\mathrm {leak}}$ & $-65$ & $\mathrm{mV}$\\

$g_{m}$ &$1.4472$ & $\mathrm{mS/cm^2}$& $g_{\mathrm {K_{SS}}}$ & $0.1512$ &$\mathrm{mS/cm^2}$\\

$g_{\mathrm {NaP}}$& $0.4307$ & $\mathrm{mS/cm^2}$ & $E_{\mathrm {NaP}}$ & $50$ & $\mathrm{mV}$\\

$g_{\mathrm {Ca}}$ & $0.54$ & $\mathrm{mS/cm^2}$ & $E_{\mathrm {Ca}}$ & $120$ & $\mathrm{mV}$\\

$I_{\mathrm {app}}$ & $8$& $\mathrm{\mu A/cm^2}$ & $C$ & $2.7$ & $\mathrm{\mu F/cm^2}$\\ 
\hline
\end{tabular}
\label{tab:para-theta}
\end{table}

\begin{table}[!htp]
\caption{Activation variable dynamics in the model \eqref{eq:main-theta}}
\centering
\begin{tabular}{|p{1cm}|p{6cm}|p{0.8cm}|p{6cm}|}
\hline
$h$ & $\tau_{\mathrm fast}=5.6115$ & $m_{\mathrm {Na}}$ & $\alpha_{m_\mathrm{Na}}(V)=-\frac{V+16}{10(\exp(-(V+16)/10)-1)}$\\
& $\alpha_h(V)=0.07\exp(-(V+30)/20)$ & &  $\beta_{m_\mathrm{Na}}(V)=4\exp(-(V+41)/18)$\\
& $\beta_h(V)=(\exp(-V/10)+1)^{-1}$ & &\\
\hline

$m_{\mathrm {K_{DR}}}$&  $\alpha_m(V)=-0.01\frac{V+20}{\exp(-(V+20)/10)-1}$ & $n$ & $n_{\infty}(V)=(1+\exp(-(V+35)/10))^{-1}$\\

& $\beta_m(V)=0.125\exp(-(V+30)/80)$ & &$\tau_n(V) = \frac{81.085}{\exp(\frac{V+35}{40})+\exp(-\frac{V+35}{20})}$ \\
\hline

$m_{\mathrm {NaP}}$& $m_\infty(V)=(1+\exp(-(V+40)/5))^{-1}$ & $s$ & $\alpha_s(V)=\frac{1.6}{(1+\exp(-0.072(V-65))}$\\
& $\tau_m=5$ & & $\beta_s(V)=0.02\frac{V-51.1}{\exp(\frac{V-51.1}{5})-1}$\\
\hline

$q$ & $\alpha_q(\mathrm{\mathrm{Ca_i}})=\mathrm {min}(0.1\mathrm{\mathrm{Ca_i}}, 1)$ & $\mathrm {\mathrm{Ca_i}}$ & $F_{\mathrm Ca}= 2.2222$ \\

& $\beta_q=0.002$ & & $\tau_{\mathrm Ca}=100$\\
\hline
\end{tabular}
\label{tab:activation}
\end{table}

\section{Dimensionless version of the cortical theta oscillator model \eqref{eq:main-theta}}\label{sec:nondim-theta}
In this section, we present the nondimensionalization of the system \eqref{eq:main-theta} to identify perturbation parameters $\varepsilon$ and $\delta$, thereby enabling the application of singular perturbation techniques. We conduct numerical simulations and find that the membrane potential typically lies between $-60$ mV and $30$ mV. Over the range $V \in [-60,30]$, we define  $T_y=\mathrm{max}\{\mathrm{max}\ (\alpha_y), \mathrm{max}\ (\beta_y)\}$ for $y \in \{ s, m_{\mathrm K_{DR}}, h,q\}$. Then, for each $y$, we define $\bar{\alpha}_y=\alpha_y/T_y$ as a rescaled version of $\alpha_y$ and $\bar{\beta}_y=\beta_y/T_y$ as a rescaled version of $\beta_y$. Furthermore, we define $\bar{\tau}_n(V)=T_n \tau_n(V)$ as a rescaled version of $\tau_n(V)$ with $T_n= \mathrm{max} (1/\tau_n(V))$. To this end, we introduce a dimensionless time variable $t_s=t/Q_t$ with a reference time scale $Q_t$, and transform \eqref{eq:main-theta} to
\begin{equation}\label{eq:nondim-theta}
\begin{array}{rcl}
    \frac{C}{g_{\mathrm {max}}Q_t}\frac{dV}{dt_s}&=& f_1(V,n,m_{\mathrm {NaP}},s,m_{\mathrm {K_{DR}}},h,q),\vspace{0.05in}\\
    \frac{\tau_m}{Q_t} \frac{dm_{\mathrm {NaP}}}{dt_s}&=& m_{\infty}(V)-m_{\mathrm NaP},\vspace{0.05in}\\
    \frac{1}{Q_tT_s}\frac{ds}{dt_s}&=& (1-s)\bar{\alpha}_s-s\bar{\beta}_s,\vspace{0.05in}\\
    \frac{1}{Q_t\tau_{\mathrm {fast}}T_{m_{\mathrm {K_{DR}}}}}\frac{dm_{\mathrm {K_{DR}}}}{dt_s}&=& (1-m_{\mathrm {K_{DR}}})\bar{\alpha}_{m}-{m_{\mathrm {K_{DR}}}}\bar{\beta}_{m},\vspace{0.05in}\\
    \frac{1}{Q_t\tau_{\mathrm {fast}}T_h}\frac{dh}{dt_s}&=& (1-h)\bar{\alpha}_h-h\bar{\beta}_h,\vspace{0.05in}\\
    \frac{1}{Q_tF_{\mathrm {Ca}}g_{\mathrm {Ca}}Q_v}\frac{d\mathrm {Ca_i}}{dt_s}&=& - \bar{I}_{\mathrm {Ca}}(V,s)-\frac{\mathrm {Ca_i}}{F_{\mathrm {Ca}}\tau_{\mathrm {Ca}}g_{\mathrm {Ca}}Q_v}, \vspace{0.05in}\\
     \frac{1}{Q_t T_n}\frac{dn}{dt_s}&=&  \frac{n_{\infty}(V)-n}{\bar{\tau}_n(V)},\vspace{0.05in}\\
     \frac{1}{Q_tT_q}\frac{dq}{dt_s}&=& (1-q)\bar{\alpha}_q( {\mathrm {Ca_i}})-q\bar{\beta}_q,
\end{array}    
\end{equation}
where $Q_v=100$ mV is a reference voltage scale, $\bar{I}_{\mathrm Ca}=I_{\mathrm {Ca}}/(g_{\mathrm {Ca}}Q_v)$ is a dimensionless calcium current, $g_{\mathrm {max}}=125$ ms is a conductance scale , and
$$f_1(V,n,m_{\mathrm {NaP}},s
,m_{\mathrm {K_{DR}}},h,q):=\frac{1}{g_{\mathrm {max}}}(I_{\mathrm {app}} - I_{\mathrm {Na}}-I_{\mathrm {K_{DR}}}-I_{\mathrm {leak}}-I_m -I_{\mathrm {NaP}}-I_{\mathrm {Ca}}-I_{\mathrm {K_{SS}}})$$
is a rescaled version of the right-hand side of the voltage equation of \eqref{eq:main-theta}. Note that it is unnecessary to rescale $V$ and $\mathrm {Ca_i}$ as they do not affect the timescales.

With that setting, numerical evaluations for $V \in [-60,30]$ show that  $T_n \approx 0.0631 \ \text{ms}^{-1}$, $\ T_s\approx 2.1620 \ \text{ms}^{-1}$,  $\ T_{m_{\mathrm {K_{DR}}}}\approx 0.5034 \ \text{ms}^{-1}$, $\ T_h \approx 0.9526 \ \text{ms}^{-1}$, $\ T_q \approx 0.0042 \ \text{ms}^{-1}$ for $\mathrm {Ca_i} \in (0,0.1)$. The parameters on the left-hand side of \eqref{eq:nondim-theta} reveal the evolution rates of the variables. We find that the voltage variable $V$ and the intracellular calcium concentration $\mathrm{Ca_i}$ have time constants $\frac{C}{g_{\mathrm {max}}} = 0.00216 \ \text{ms} = O(0.001)$ and $ \frac{1}{F_{\mathrm {Ca}}g_{\mathrm {Ca}}Q_v}= 0.0083 \ \text{ms}=O(0.001)$, respectively. The activation variables $m_{\mathrm{Nap}}$, $s$, $m_\mathrm{K_{DR}}$, $h$ operate on slower timescales with $\tau_m= 3 \ \text{ms}= O(1)$, $ \frac{1}{T_s}\approx 0.4625 \ \text{ms}=O(0.1)$, $\frac{1}{\tau_{\mathrm {fast}}T_{m_{\mathrm {K_{DR}}}}}\approx 0.3337 \ \text{ms}=O(0.1)$, and $\frac{1}{\tau_{\mathrm {fast}}T_h}\approx 0.1871 \ \text{ms} = O(0.1)$, respectively. However, they evolve faster than the gating variable $n$ with a timescale of $ \frac{1}{T_n}\approx 15.84 \ \text{ms}=O(10)$ and are substantially faster than $q$ with the time constant $ \frac{1}{T_q}\approx 238.0952 \ \text{ms}=O(100)$. For simplicity, we therefore group $(V, m_{\mathrm{Nap}}, s, m_\mathrm{K_{DR}}, h, \mathrm{Ca_i})$ as fast variables, classify $n$ as a slow variable and $q$ as a superslow variable.

We pick $Q_t=1/T_n$ and define the small parameters as follows
$$\bar{\varepsilon}_1:=\frac{C}{g_{\mathrm {max}}Q_t}, \quad \bar{\varepsilon}_2:=\frac{\tau_m}{Q_t}, \quad \bar{\varepsilon}_3:=\frac{1}{Q_tT_s},\quad \bar{\varepsilon}_4:=\frac{1}{Q_t\tau_{\mathrm {fast}}T_{m_{\mathrm {K_{DR}}}}},$$
$$\bar{\varepsilon}_5:=\frac{1}{Q_t\tau_{\mathrm {fast}}T_h}, \quad \bar{\varepsilon}_6:=\frac{1}{Q_tF_{\mathrm {Ca}}g_{\mathrm {Ca}}Q_v}, \quad \bar{\delta}:=Q_t T_q.$$ 
Consequently, the system \eqref{eq:nondim-theta} becomes
\begin{equation}\label{eq:nondim-theta-2}
\begin{array}{rcl}
    \bar{\varepsilon}_1\frac{dV}{dt_s}&=& f_1(V,n,m_{\mathrm {NaP}},s,m_{\mathrm {K_{DR}}},h,q),\vspace{0.05in}\\
    \bar{\varepsilon}_2 \frac{dm_{\mathrm {NaP}}}{dt_s}&=& m_{\infty}(V)-m_{\mathrm NaP}:=f_2(V,m_\mathrm{NaP}),\vspace{0.05in}\\
   \bar{\varepsilon}_3\frac{ds}{dt_s}&=& (1-s)\bar{\alpha}_s-s\bar{\beta}_s := f_3(V,s),\vspace{0.05in}\\
   \bar{\varepsilon}_4 \frac{dm_{\mathrm {K_{DR}}}}{dt_s}&=& (1-m_{\mathrm {K_{DR}}})\bar{\alpha}_{m}-{m_{\mathrm {K_{DR}}}}\bar{\beta}_{m}:= f_4(V, m_\mathrm{K_{DR}}),\vspace{0.05in}\\
    \bar{\varepsilon}_5\frac{dh}{dt_s}&=& (1-h)\bar{\alpha}_h-h\bar{\beta}_h := f_5(V,h),\vspace{0.05in}\\
    \bar{\varepsilon}_6 \frac{d\mathrm {Ca_i}}{dt_s}&=& - \bar{I}_{\mathrm {Ca}}(V,s)-\frac{\mathrm {Ca_i}}{F_{\mathrm {Ca}}\tau_{\mathrm {Ca}}g_{\mathrm {Ca}}Q_v} := f_6(V,s,\mathrm{Ca_i}), \vspace{0.05in}\\
    \frac{dn}{dt_s}&=&  \frac{n_{\infty}(V)-n}{\bar{\tau}_n(V)} := g_1(V,n),\vspace{0.05in}\\
    \frac{dq}{dt_s}&=&\bar{\delta}((1-q)\bar{\alpha}_q( {\mathrm {Ca_i}})-q\bar{\beta}_q):=\bar{\delta} g_2(\mathrm{Ca_i},q),
\end{array}    
\end{equation}
where 
$\bar{\varepsilon}_1 \approx 0.0014$, $\bar{\varepsilon}_2 \approx 0.3230$, $\bar{\varepsilon}_3 \approx 0.0292$, $\bar{\varepsilon}_4 \approx 0.0211$, $\bar{\varepsilon}_5 \approx 0.0118$, $\bar{\varepsilon}_6 \approx 0.0005$, and $\bar{\delta}\approx 0.0666$. 

The distinct values of $\bar{\varepsilon_i}$ indicate that $V$ and the variables $(m_{\mathrm{Nap}}, s, m_\mathrm{K_{DR}}, h, \mathrm{\mathrm{Ca_i}})$ do not operate on the same timescale. Nevertheless, their timescales remain well separated from those of $n$
and $q$, justifying their collective treatment as fast variables. Thus, system \eqref{eq:nondim-theta-2} admits a three-timescale singular perturbation structure.

To complement the GSPT analysis, we introduce scaling factors $r_\varepsilon$ and $r_\delta$ to vary the fast and superslow timescales. Specifically, we multiply the right-hand sides of the fast and superslow equations by $r_\varepsilon$ and $r_\delta$, respectively, yielding
\begin{equation}\label{eq:nondim-theta-3}
\begin{array}{rcl}
    \bar\varepsilon_1\frac{dV}{dt_s}&=& r_\varepsilon f_1(V,n,m_{\mathrm {NaP}},s,m_{\mathrm {K_{DR}}},h,q),\vspace{0.05in}\\
    \bar\varepsilon_2 \frac{dm_{\mathrm {NaP}}}{dt_s}&=& r_\varepsilon f_2(V,m_\mathrm{NaP}),\vspace{0.05in}\\
   \bar\varepsilon_3\frac{ds}{dt_s}&=&r_\varepsilon f_3(V,s),\vspace{0.05in}\\
  \bar\varepsilon_4 \frac{dm_{\mathrm {K_{DR}}}}{dt_s}&=& r_\varepsilon f_4(V, m_\mathrm{K_{DR}}),\vspace{0.05in}\\
    \bar\varepsilon_5\frac{dh}{dt_s}&=& r_\varepsilon f_5(V,h),\vspace{0.05in}\\
    \bar\varepsilon_6 \frac{d\mathrm {Ca_i}}{dt_s}&=&  r_\varepsilon f_6(V,s,\mathrm{Ca_i}), \vspace{0.05in}\\
    \frac{dn}{dt_s}&=&  g_1(V,n),\vspace{0.05in}\\
    \frac{dq}{dt_s}&=& r_\delta\bar\delta  g_2(\mathrm{Ca_i},q).
\end{array}    
\end{equation}
We define the rescaled small perturbation parameters as follows
$$\varepsilon_1 := \frac{\bar{\varepsilon}_1}{r_\varepsilon}, \quad \varepsilon_2 := \frac{\bar{\varepsilon}_2}{r_\varepsilon}, \quad \varepsilon_3 := \frac{\bar{\varepsilon}_3}{r_\varepsilon}, \quad \varepsilon_4 := \frac{\bar{\varepsilon}_4}{r_\varepsilon},\quad \varepsilon_5 := \frac{\bar{\varepsilon}_5}{r_\varepsilon}, \quad \varepsilon_6 := \frac{\bar{\varepsilon}_6}{r_\varepsilon}, \quad \delta:=r_\delta\bar{\delta},$$
which are identical to the original perturbation parameters in the default case when $r_\varepsilon=r_\delta=1$.

With this rescaling, system \eqref{eq:nondim-theta-3} can be rewritten as 
\begin{equation} \label{eq:slow-sys-appendix}
\begin{array}{rcl}
    \varepsilon_1\frac{dV}{dt_s}&=& f_1(V,n,m_{\mathrm {NaP}},s,m_{\mathrm {K_{DR}}},h,q),\vspace{0.05in}\\
    \varepsilon_2 \frac{dm_{\mathrm {NaP}}}{dt_s}&=&  f_2(V,m_\mathrm{NaP}),\vspace{0.05in}\\
   \varepsilon_3\frac{ds}{dt_s}&=&f_3(V,s),\vspace{0.05in}\\
   \varepsilon_4 \frac{dm_{\mathrm {K_{DR}}}}{dt_s}&=&  f_4(V, m_\mathrm{K_{DR}}),\vspace{0.05in}\\
    \varepsilon_5\frac{dh}{dt_s}&=&  f_5(V,h),\vspace{0.05in}\\
    \varepsilon_6 \frac{d\mathrm {Ca_i}}{dt_s}&=&   f_6(V,s,\mathrm{Ca_i}), \vspace{0.05in}\\
    \frac{dn}{dt_s}&=&  g_1(V,n),\vspace{0.05in}\\
    \frac{dq}{dt_s}&=& \delta  g_2(\mathrm{Ca_i},q),
\end{array}    
\end{equation}
which is the slow system \eqref{eq:slow-theta} presented in Section \ref{sec:model}.

The relationship between the default and rescaled perturbation parameters, along with their default values, is summarized in Table~\ref{tab:eps-delta-para}. Increasing $r_\varepsilon$ decreases all $\varepsilon_i$, thereby speeding up all fast variables, whereas decreasing $r_\delta$ reduces $\delta$ and hence slows down the superslow variable.

\begin{table}[ht]
\centering
\caption{Rescaled perturbation parameters. The original (default) parameters $\bar{\varepsilon}_i$ are rescaled by $r_\varepsilon$, and $\bar{\delta}$ by $r_\delta$, to vary timescales.}
\label{tab:eps-delta-para}
\begin{tabular}{|c|cccccc|c|}
\hline
Default Parameter & $\bar{\varepsilon}_1$ & $\bar{\varepsilon}_2$ & $\bar{\varepsilon}_3$ & $\bar{\varepsilon}_4$ & $\bar{\varepsilon}_5$ & $\bar{\varepsilon}_6$ & $\bar{\delta}$ \\
\hline
Rescaled Parameter & 
\multicolumn{6}{c|}{\rule[-2.5ex]{0pt}{6ex} $\varepsilon_i = \dfrac{\bar{\varepsilon}_i}{r_\varepsilon}, \quad i=1,\dots,6$} &
$\delta= r_\delta \bar{\delta}$ \\
\hline
Default Value & $0.0014$ & $0.3230$ & $0.0292$ & $0.0211$ & $0.0118$ & $0.0005$ & $0.0666$ \\
\hline
\end{tabular}
\end{table}

\section{Geometric Singular Perturbation Analysis and Singular Limits}\label{ap:singular-limit}

In this section, we perform GSPT analysis on system \eqref{eq:slow-sys-appendix}, which is described over the slow timescale $t_s$. We call \eqref{eq:slow-sys-appendix} the \textit{slow system}. Equivalent descriptions of the dynamics can be obtained through appropriate time rescalings. 

Defining a fast time $t_f=t_s/\varepsilon_2$ yields the \textit{fast system}:
\begin{equation}\label{eq:fast-theta}
\begin{array}{rcl}
   \frac{dV}{dt_f}&=& \frac{\varepsilon_2}{\varepsilon_1}  f_1(V,n,m_{\mathrm {NaP}},s,m_{\mathrm {K_{DR}}},h,q),\vspace{0.05in}\\
    \frac{dm_{\mathrm {NaP}}}{dt_f}&=&  f_2(V,m_\mathrm{NaP}),\vspace{0.05in}\\
   \frac{ds}{dt_f}&=& \frac{\varepsilon_2}{\varepsilon_3}  f_3(V,s),\vspace{0.05in}\\
   \frac{dm_{\mathrm {K_{DR}}}}{dt_f}&=& \frac{\varepsilon_2}{\varepsilon_4}  f_4(V, m_\mathrm{K_{DR}}),\vspace{0.05in}\\
    \frac{dh}{dt_f}&=& \frac{\varepsilon_2}{\varepsilon_5}  f_5(V,h),\vspace{0.05in}\\
     \frac{d\mathrm {\mathrm{Ca_i}}}{dt_f}&=& \frac{\varepsilon_2}{\varepsilon_6}  f_6(V,s,\mathrm{\mathrm{Ca_i}}), \vspace{0.05in}\\
    \frac{dn}{dt_f}&=& \varepsilon_2  g_1(V,n),\vspace{0.05in}\\
    \frac{dq}{dt_f}&=& \varepsilon_2 \delta  g_2(\mathrm{\mathrm{Ca_i}},q),
\end{array}      
\end{equation}
which evolves on the \textit{fast timescale}. 
Alternatively, introducing a superslow time $t_{ss}=\delta t_s$, we obtain the  \textit{superslow system}:
\begin{equation}\label{eq:superslow-theta}
\begin{array}{rcl}
    \varepsilon_1 \delta \frac{dV}{dt_{ss}}&=&  f_1(V,n,m_{\mathrm {NaP}},s,m_{\mathrm {K_{DR}}},h,q),\vspace{0.05in}\\
    \varepsilon_2 \delta \frac{dm_{\mathrm {NaP}}}{dt_{ss}}&=&  f_2(V,m_\mathrm{NaP}),\vspace{0.05in}\\
    \varepsilon_3 \delta \frac{ds}{dt_{ss}}&=& f_3(V,s),\vspace{0.05in}\\
   \varepsilon_4 \delta \frac{dm_{\mathrm {K_{DR}}}}{dt_{ss}}&=&  f_4(V, m_\mathrm{K_{DR}}),\vspace{0.05in}\\
  \varepsilon_5 \delta \frac{dh}{dt_s}&=&  f_5(V,h),\vspace{0.05in}\\
   \varepsilon_6 \delta \frac{d\mathrm {\mathrm{Ca_i}}}{dt_{ss}}&=& f_6(V,s,\mathrm{\mathrm{Ca_i}}), \vspace{0.05in}\\
 \delta   \frac{dn}{dt_{ss}}&=&  g_1(V,n),\vspace{0.05in}\\
    \frac{dq}{dt_{ss}}&=&  g_2(\mathrm{\mathrm{Ca_i}},q),
\end{array}    
\end{equation}
which evolves on the \textit{superslow timescale}. 

For convenience, we use $\varepsilon$ to denote the collection $(\varepsilon_1, \cdots, \varepsilon_6)$ unless specified otherwise. 
In the following subsections, we apply GSPT from three perspectives: first, by treating $\varepsilon$ as the only singular perturbation parameters (Section \ref{sub:eps-viewpoint-theta}); second, by treating $\delta$ as the only perturbation parameter (Section \ref{sub:delta-viewpoint-theta}), and finally, by treating $\varepsilon$ and $\delta$ as independent singular perturbation parameters (Section \ref{sec:double-singular-theta}). 

\subsection{$\varepsilon \rightarrow 0$ singular limit and canard dynamics}\label{sub:eps-viewpoint-theta}

In this subsection, we fix $\delta>0$ and consider the singular limit $\varepsilon \rightarrow 0$ to derive the fast layer problem and the slow reduced problem. Since the system involves six perturbation parameters ($\varepsilon_1,\cdots,\varepsilon_6$), associated with six fast variables, it is necessary to specify how this limit is taken. 

To this end, we assume that the perturbation parameters vanish proportionally, in the sense that 
\[
\lim_{(\varepsilon_j,\ \varepsilon_2) \rightarrow (0, 0)} \frac{\varepsilon_2}{\varepsilon_j}=k_j, \ \text{or} \ \varepsilon_2=k_j\varepsilon_j \ \text{for} \ j= 1, 3, 4, 5, 6,
\]
where $k_j>0$ are constants. This allows for differences in the relative evolving rates of the six fast variables, while ensuring that they remain on a common fast timescale relative to the slow variables. This limiting can be realized through the scaling factor $r_\varepsilon$ introduced in \eqref{eq:nondim-theta-3}: increasing $r_\varepsilon$ decreases all $\varepsilon_i$ proportionally, thereby driving the system toward the singular limit $(\varepsilon_1,\cdots,\varepsilon_6)\rightarrow 0$.

Now taking $\varepsilon_2 \rightarrow 0$ (and thus $(\varepsilon_1, \varepsilon_3, \varepsilon_4, \varepsilon_5, \varepsilon_6) \rightarrow (0,0,0,0,0)$) in the fast system \eqref{eq:fast-theta} yields the six-dimensional
(6D) \textit{fast layer problem}:
\begin{equation}\label{eq:fast-subsystem-theta}
\begin{array}{rcl}
   \frac{dV}{dt_f}&=& k_1  f_1(V,n,m_{\mathrm {NaP}},s,m_{\mathrm {K_{DR}}},h,q),\vspace{0.05in}\\
    \frac{dm_{\mathrm {NaP}}}{dt_f}&=&  f_2(V,m_\mathrm{NaP}),\vspace{0.05in}\\
   \frac{ds}{dt_f}&=& k_3  f_3(V,s),\vspace{0.05in}\\
   \frac{dm_{\mathrm {K_{DR}}}}{dt_f}&=& k_4  f_4(V, m_\mathrm{K_{DR}}),\vspace{0.05in}\\
    \frac{dh}{dt_f}&=& k_5 f_5(V,h),\vspace{0.05in}\\
     \frac{d\mathrm {\mathrm{Ca_i}}}{dt_f}&=& k_6 f_6(V,s,\mathrm{\mathrm{Ca_i}}), 
\end{array}      
\end{equation}
which describes the dynamics of the fast variables for fixed values of the other variables.

The set of equilibrium points of the fast layer problem \eqref{eq:fast-subsystem-theta} is called the \textit{critical manifold}, which is given by   
\begin{equation}
M_{s}:=\{(V, m_{\mathrm {NaP}}, s, m_{\mathrm {K_{DR}}}, h, \mathrm{\mathrm{Ca_i}}, n, q):\ f_1 = f_2 = f_3 = f_4 = f_5 = f_6 = 0\}.
\end{equation}
Since the variables $m_{\mathrm {NaP}}$, $s$, $m_{\mathrm {K_{DR}}}$, $h$, $\mathrm{\mathrm{Ca_i}}$, and $q$ enter linearly in $f_1, \cdots, f_6$, the equations $f_1=\cdots=f_6=0$ can be solved explicitly for these variables in terms $V$ and $n$. Thus, the two-dimensional critical manifold $M_s$ can be represented as a graph in terms of $(V, n)$ as follows:
\begin{equation} \label{eq:ms-representation-theta}
    \begin{array}{rcl}
        m_{\mathrm {NaP}} &=& m_{\infty}(V), \vspace{0.05in}\\
        s &=& \frac{\alpha_{s}(V)}{\alpha_{s}(V) + \beta_{s}(V)}, \vspace{0.05in}\\
        m_\mathrm{K_{DR}} &=& \frac{\alpha_{m}(V)}{\alpha_{m}(V) + \beta_{m}(V)}, \vspace{0.05in}\\
        h &=& \frac{\alpha_{h}(V)}{\alpha_{h}(V) + \beta_{h}(V)}, \vspace{0.05in}\\
       \mathrm{\mathrm{Ca_i}} &=& -F_{\mathrm{Ca}}I_{Ca}\tau_{\mathrm{Ca}},\\
        q &=& \frac{I_{\mathrm {app}} - I_{\mathrm {Na}} - I_{\mathrm {K_{DR}}} - I_{\mathrm {leak}} - I_{m} - I_{\mathrm {NaP}} - I_{\mathrm Ca}}{g_{\mathrm {K_{SS}}} (V-E_{\mathrm K})}. \vspace{0.05in}
    \end{array}
\end{equation}

$M_s$ is folded along a set of saddle-node bifurcations of the fast layer problem \eqref{eq:fast-subsystem-theta}, defined by:
\begin{equation}\label{eq:fold-theta}
    L_s:=\{ (V, m_{\mathrm {NaP}}, s, m_{\mathrm {K_{DR}}}, h, \mathrm{\mathrm{Ca_i}}, n, q) \in M_s: \det(\mathbf{J}_{k_j})=0 \},
\end{equation}
where $\mathbf{J}_{k_j}$ denotes the $6 \times 6$ Jacobian matrix of the fast layer subsystem, and $\det(\mathbf{J}_{k_j})$ denotes the determinant of $\mathbf{J}_{k_j}$. The Jacobian matrix $\mathbf{J}_{k_j}$ is given explicitly by:
\begin{equation}\label{eq:J-matrix-theta}
\mathbf{J}_{k_j}=
\begin{pmatrix}
k_1  f_{1V} & k_1 f_{1m_{\mathrm {NaP}}} & k_1  f_{1s} & k_1  f_{1m_{\mathrm{K_{DR}}}}& k_1  f_{1h} & 0  \\
 f_{2V} &  f_{2m_{\mathrm {NaP}}} & 0 & 0 & 0 & 0 \\
k_3  f_{3V} & 0 & k_3  f_{3s} & 0 & 0 & 0 \\
k_4  f_{4V} & 0 & 0 & k_4  f_{4m_{\mathrm {K_{DR}}}} & 0 & 0  \\
k_5  f_{5V} & 0 & 0 & 0 & k_5  f_{5h} & 0 \\
k_6  f_{6V} & 0 & k_6  f_{6s} & 0 & 0 & k_6 f_{6 \mathrm{\mathrm{Ca_i}}}  \\
\end{pmatrix}.
\end{equation}
A direct computation of the determinant yields:
 \begin{equation} \label{eq:det-J_theta}
 \begin{array}{rcl}
\frac{1}{ k_1 k_3 k_4 k_5 k_6}\det(\mathbf{J}_{k_j}) &=& f_{1V}f_{2m_{\mathrm{NaP}}}f_{3s}f_{4m_{\mathrm{K_{DR}}}}f_{5h}f_{6\mathrm{\mathrm{Ca_i}}} \\
&& - f_{1m_{\mathrm{NaP}}}f_{2V}f_{3s}f_{4m_{\mathrm{K_{DR}}}}f_{5h}f_{6\mathrm{\mathrm{Ca_i}}}\\
&&- f_{1s}f_{2m_{\mathrm{NaP}}}f_{3V}f_{4m_{\mathrm{K_{DR}}}}f_{5h}f_{6\mathrm{\mathrm{Ca_i}}}\\
&&-f_{1m_{\mathrm{K_{DR}}}}f_{2m_{\mathrm{NaP}}}f_{3s}f_{4V}f_{5h}f_{6\mathrm{\mathrm{Ca_i}}}\\
&&-f_{1h}f_{2m_{\mathrm{NaP}}}f_{3s}f_{4m_{\mathrm{K_{DR}}}}f_{5V}f_{6\mathrm{\mathrm{Ca_i}}}.
\end{array}
\end{equation}
Here, $f_{jx}$ denotes the partial derivatives of the function $f_j$ for $j=1, \dots, 6$ with respect to the variable $x$ for $x \in \{V, m_{\mathrm {NaP}}, s, m_{\mathrm {K_{DR}}}, h, \mathrm{\mathrm{Ca_i}}\}$. 
There are two fold curves that separate the critical manifold $M_s$ into three sheets: a lower attracting sheet $M_s^L$ and middle and upper repelling sheets $M_s^M$ and $M_s^U$.

Taking the same limit $\varepsilon \rightarrow 0$ with $\delta >0$ in the slow system \ref{eq:slow-sys-appendix} yields the 2D \textit{slow reduced problem}:
\begin{equation}\label{eq:slow-reduced-theta}
\begin{array}{rcl}
    \frac{dn}{dt_s}&=&  g_1(V,n),\vspace{0.05in}\\
    \frac{dq}{dt_s}&=& \delta  g_2(\mathrm{\mathrm{Ca_i}},q),
\end{array} 
\end{equation}
that describes the slow flow on the critical manifold $M_s$. For sufficiently small $\varepsilon$, the slow motions along $M_s$ perturb smoothly to the flow of \eqref{eq:slow-sys-appendix} restricted to locally invariant slow manifolds for sufficiently small $\varepsilon$ \cite{Fenichel1979}.

To investigate the dynamics of \eqref{eq:slow-reduced-theta} on $M_s$, we differentiate the algebraic constraints \eqref{eq:slow-reduced-theta} and rearrange terms appropriately to obtain:
\begin{equation}\label{eq:algebraic-theta}
\begin{pmatrix}
-\mathbf{J} & \mathbf{0} \\ \mathbf{0} & \mathbf{I_2}
\end{pmatrix} 
\frac{d}{dt_s} 
\begin{pmatrix}
V \\ m_{\mathrm{ NaP}} \\ s \\ m_{\mathrm {K_{DR}}} \\ h \\ \mathrm{\mathrm{Ca_i}} \\ n \\q
\end{pmatrix} =
\begin{pmatrix}
 f_{1n}  g_1 + f_{1q}\delta  g_2  \\
0\\
0 \\
0\\
0\\0\\
 g_1\\
\delta  g_2
\end{pmatrix},
\end{equation}
where $\mathbf{J}$ is a special case of the Jacobian matrix $\mathbf{J}_{k_j}$ with $k_j=1$ for $j= 1, 3, 4, 5, 6$, $\mathbf{I_2}$ is the $2 \times 2$ identity matrix, $\mathbf{0}$ is the $6 \times 6$ zero matrix, $f_{1n}$ and $f_{1q}$ are partial derivatives with respect to $n$ and $q$, respectively. We then multiply both sides of \eqref{eq:algebraic-theta} by $\begin{pmatrix}
    - \mathrm{adj} (\mathbf{J}) & \mathbf{0} \\ \mathbf{0} & \mathbf{I_2}
\end{pmatrix}$ to obtain the dynamics of \eqref{eq:slow-reduced-theta} in all coordinate charts:
\begin{equation}\label{eq:adj-J-theta}
\begin{pmatrix}
\det(\mathbf{J}) \mathbf{I_6} & \mathbf{0} \\ \mathbf{0} & \mathbf{I_2}
\end{pmatrix} 
\frac{d}{dt_s} 
\begin{pmatrix}
V \\ m_{\mathrm {NaP}} \\ s \\ m_{\mathrm {K_{DR}}} \\ h \\ \mathrm{\mathrm{Ca_i}} \\ n \\q
\end{pmatrix} = \begin{pmatrix}
- \mathrm{adj} (\mathbf{J}) & \mathbf{0} \\ \mathbf{0} & \mathbf{I_2}
\end{pmatrix} 
\begin{pmatrix}
 f_{1n}  g_1 + f_{1q}\delta  g_2  \\
0\\
0 \\
0\\
0\\0\\
 g_1\\
\delta  g_2
\end{pmatrix}.
\end{equation}
Here, $\det(\mathbf{J})$ is the determinant of $\mathbf{J}$, $\mathrm{adj} (\mathbf{J})$ is the adjoint of $\mathbf{J}$, $\mathbf{I_2}$ and $\mathbf{I_6}$ are the $2 \times 2$ and $6 \times 6$ identity matrices, respectively. Through direct computations, the right-hand side of \eqref{eq:adj-J-theta} is given by:
\begin{equation}
\begin{pmatrix}
-  f_{2m_{\mathrm {NaP}}} f_{3s} f_{4m_{\mathrm {K_{DR}}}} f_{5h} f_{6\mathrm{Ca_i}} \left( f_{1n}  g_1 + f_{1q}\delta  g_2 \right) \\
 f_{2V} f_{3s} f_{4m_{\mathrm {K_{DR}}}} f_{5h} f_{6\mathrm{Ca_i}} \left(f_{1n}  g_1 + f_{1q}\delta  g_2 \right)\\
 f_{3V} f_{2m_{\mathrm {NaP}}} f_{4m_{\mathrm {K_{DR}}}} f_{5h} f_{6\mathrm{Ca_i}} \left( f_{1n}  g_1 + f_{1q}\delta  g_2 \right) \\
 f_{4V} f_{2m_{\mathrm {NaP}}} f_{3s} f_{5h} f_{6\mathrm{Ca_i}} \left( f_{1n}  g_1 + f_{1q}\delta  g_2 \right)\\
 f_{5V} f_{2m_{\mathrm {NaP}}} f_{3s} f_{4m_{\mathrm {K_{DR}}}} f_{6\mathrm{Ca_i}} \left( f_{1n}  g_1 + f_{1q}\delta  g_2 \right)\\
 (-f_{3V} f_{2m_{\mathrm {NaP}}}f_{6s}f_{4m_{\mathrm {K_{DR}}}} f_{5h} +
f_{6V} f_{2m_{\mathrm {NaP}}} f_{3s} f_{4m_{\mathrm {K_{DR}}}} f_{5h}) \left( f_{1n}  g_1 + f_{1q}\delta  g_2 \right)\\
 g_1\\
\delta  g_2
\end{pmatrix}.
\end{equation}
Since the critical manifold has the graph representation \eqref{eq:ms-representation-theta}, we can obtain the projection of \eqref{eq:slow-reduced-theta} on the $(V,n)$-plane as following:
\begin{equation} \label{eq:proj-dlow-reduced-theta}
    \begin{array}{rcl}
        \det(\mathbf{J}) \frac{dV}{dt_{s}} &=& -  f_{2m_{\mathrm {NaP}}} f_{3s} f_{4m_{\mathrm {K_{DR}}}} f_{5h} f_{6\mathrm{Ca_i}} \left( f_{1n}  g_1 + f_{1q}\delta g_2 \right) , \vspace{0.05in}\\
        \frac{dn}{dt_{s}} &=&  g_1.
    \end{array}
\end{equation}
We note that \eqref{eq:proj-dlow-reduced-theta}
is singular at the fold curves \eqref{eq:fold-theta}, i.e., $\det(\mathbf{J})=0$. However, this singular term can be removed through coordinate transformation,  $t_s=\det(\mathbf{J})t_d$, to give the desingularized system:
\begin{equation}\label{eq:desingularization-theta}
    \begin{array}{rcl}
        \frac{dV}{dt_{d}} &=&  -  f_{2m_{\mathrm {NaP}}} f_{3s} f_{4m_{\mathrm {K_{DR}}}} f_{5h} f_{6\mathrm{Ca_i}} \left( f_{1n}  g_1 + f_{1q}\delta  g_2 \right) \vspace{0.05in}\\
        &:=& F_\delta (V, m_{\mathrm {NaP}}, s, m_{\mathrm {K_{DR}}}, h, \mathrm{Ca_i}, n,q, \delta)  \vspace{0.05in} \\
        \frac{dn}{dt_{d}} &=& \det(\mathbf{J})  g_1.
    \end{array}
\end{equation}

The desingularized system \eqref{eq:desingularization-theta} has the same phase portrait as the system \eqref{eq:proj-dlow-reduced-theta}, i.e., when $\det(\mathbf{J})>0$. However, when $\det(\mathbf{J})<0$, the time transformation reverses the orientation of trajectories of the system \eqref{eq:desingularization-theta} to obtain the phase portrait of the system \eqref{eq:proj-dlow-reduced-theta}.
There are two kinds of singularities in the desingularized system \eqref{eq:desingularization-theta}: ordinary and folded. The ordinary singularities are the true equilibria of the full-system \eqref{eq:slow-theta}, given by:
\begin{equation} \label{eq:ordinary-theta}
    \mathrm{TE}:=\{ (V, m_{\mathrm {NaP}}, s, m_{\mathrm {K_{DR}}}, h, \mathrm{Ca_i}, n, q) \in M_s: g_1(V,n) = g_2(\mathrm{Ca_i}, q) = 0\}.
\end{equation}
For the default parameters, TE lies on the middle repelling sheet $M_s^M$ of $M_s$ and is a saddle-focus equilibrium. Folded singularities, on the other hand, are not equilibria of the full-system \eqref{eq:slow-theta}. They are points on the fold curves $L_s$ at which the right-hand side of the $V$-equation in the desingularized system \eqref{eq:desingularization-theta} vanishes. Since $f_{2m_{\mathrm {NaP}}}$, $f_{3s}$, $f_{4m_{\mathrm {K_{DR}}}}$, $f_{5h}$, and $f_{6\mathrm{Ca_i}}$ are nonzero, folded singularities are defined as:
\begin{equation} \label{eq:folded-singu-orig-theta}
\begin{array}{rcl}
    \mathcal{M_\delta}&:=&\{ (V, m_{\mathrm {NaP}}, s, m_{\mathrm {K_{DR}}}, h, \mathrm{Ca_i}, n, q) \in L_s:  F_\delta (V, m_{\mathrm {NaP}}, s, m_{\mathrm {K_{DR}}}, h, \mathrm{Ca_i}, n,q, \delta) = 0 \} \vspace{0.05in}\\
    &=& \{ (V, m_{\mathrm {NaP}}, s, m_{\mathrm {K_{DR}}}, h, \mathrm{Ca_i}, n, q) \in L_s: g_1=\delta \frac{f_{1q}g_2}{f_{1n}}\}.
    \end{array}
\end{equation}

Folded singularities are special points that make both sides of the $V$-equation in the system \eqref{eq:proj-dlow-reduced-theta} disappear. That is,  there is a L'H\^{o}pital-type limit at a folded singularity, which allows the trajectory to pass through the folds $L_s$ with finite speed. Such trajectories are called singular canards \cite{SW2001, Wechselberger2005, Wechselberger2012}, and folded singularities contain the key information needed to understand canard dynamics under small perturbations ($0 < \varepsilon \ll 1$). 
Based on the linearization of the desingularized system \eqref{eq:desingularization-theta} at the folded singularity, there are three generic folded singularity types: \textit{folded saddles} (with real eigenvalues of the opposite signs), \textit{folded nodes} (with real eigenvalues of the same sign), and \textit{folded foci} (with complex eigenvalues). The boundary between a folded node and a folded saddle is called a \textit{folded saddle node}.


The condition \eqref{eq:folded-singu-orig-theta} implies that $\mathcal{M_\delta}$ lies $O(\delta)$ close to the intersection point of the fold curves $L_s$ and the set, $\{ (V, m_{\mathrm {NaP}}, s, m_{\mathrm {K_{DR}}}, h, \mathrm{Ca_i}, n, q) \in M_s: g_1 = 0 \}$, which is the superslow manifold (see Section \ref{sub:delta-viewpoint-theta}). This intersection point is the so-called canard-delayed-Hopf (CDH) singularity \cite{Vo2013, Letson2017, Phan2023}. 
Fig. \ref{fig:twopar-hopf-theta}A shows that, as $r_\delta$ decreases from $10$, the folded singularity $\mathcal{M_\delta}$ transitions from a folded focus (dashed blue) to a folded node (solid blue), which then converges to the CDH singularity and becomes a folded saddle-node as $r_\delta \rightarrow 0$ (i.e., $\delta \rightarrow 0$). For $r_\delta<0$, the folded singularity becomes a folded saddle. Since we only consider $r_\delta\geq 0$, this regime is not shown in panel A. 


\subsection{$\delta \rightarrow 0$ singular limit and the delayed Hopf bifurcation (DHB)}\label{sub:delta-viewpoint-theta}

Letting $\delta \rightarrow 0$ with fixed $\varepsilon_1, \varepsilon_2, \varepsilon_3, \varepsilon_4, \varepsilon_5, \varepsilon_6$ (collectively denoted as $\varepsilon$) in the slow system \eqref{eq:slow-theta} gives the 7D \textit{slow layer problem}:
\begin{equation}\label{eq:slow-layer-theta}
\begin{array}{rcl}
    \varepsilon_1\frac{dV}{dt_s}&=&  f_1(V,n,m_{\mathrm {NaP}},s,m_{\mathrm {K_{DR}}},h,q),\vspace{0.05in}\\
    \varepsilon_2 \frac{dm_{\mathrm {NaP}}}{dt_s}&=&  f_2(V,m_\mathrm{NaP}),\vspace{0.05in}\\
   \varepsilon_3\frac{ds}{dt_s}&=& f_3(V,s),\vspace{0.05in}\\
   \varepsilon_4 \frac{dm_{\mathrm {K_{DR}}}}{dt_s}&=&  f_4(V, m_\mathrm{K_{DR}}),\vspace{0.05in}\\
    \varepsilon_5\frac{dh}{dt_s}&=&  f_5(V,h),\vspace{0.05in}\\
    \varepsilon_6 \frac{d\mathrm {\mathrm{Ca_i}}}{dt_s}&=&   f_6(V,s,\mathrm{\mathrm{Ca_i}}), \vspace{0.05in}\\
    \frac{dn}{dt_s}&=&  g_1(V,n),
\end{array}    
\end{equation}
where the superslow variable $q$ is a parameter. The critical manifold $M_{ss}$ of the slow layer problem  contains the equilibria of \eqref{eq:slow-layer-theta} and is defined as a 1D subset of $M_s$ as follows:
\begin{equation}
M_{ss}:=\{(V, m_{\mathrm {NaP}}, s, m_{\mathrm {K_{DR}}}, h, \mathrm{Ca_i}, n, q): 
f_1 = f_2 = f_3 = f_4 = f_5 = f_6 = g_1 = 0\}.
\end{equation}

Similar to $M_s$, we derive an explicit condition for isolated saddle-node bifurcations of the slow layer problem \eqref{eq:slow-layer-theta}, at which GSPT \cite{Fenichel1979} breaks down, as follows:
\begin{equation}\label{eq:fold-Mss-theta}
    \mathcal{L}:=\{ (V, m_{\mathrm {NaP}}, s, m_{\mathrm {K_{DR}}}, h, \mathrm{\mathrm{Ca_i}}, n, q) \in M_s: \det(J_{SL})=0 \},
\end{equation}
where $J_{SL}$ is the $7 \times 7$ Jacobian matrix of the slow layer problem \eqref{eq:slow-layer-theta} given by:
\begin{equation} \label{eq:J-SL_theta}
J_{SL}=
\begin{pmatrix}
\frac{1}{\varepsilon_1}f_{1V} & \frac{1}{\varepsilon_1}f_{1m_{\mathrm {NaP}}} & \frac{1}{\varepsilon_1}f_{1s} & \frac{1}{\varepsilon_1}f_{1m_{\mathrm {K_{DR}}}}& \frac{1}{\varepsilon_1}f_{1h} & 0 &   \frac{1}{\varepsilon_1}f_{1n} \\
\frac{1}{\varepsilon_2}f_{2V} & \frac{1}{\varepsilon_2}f_{2m_{\mathrm {NaP}}} & 0 & 0 & 0 & 0 & 0\\
\frac{1}{\varepsilon_3}f_{3V} & 0 & \frac{1}{\varepsilon_3}f_{3s} & 0 & 0 & 0 & 0\\
\frac{1}{\varepsilon_4}f_{4V} & 0 & 0 & \frac{1}{\varepsilon_4}f_{4m_{\mathrm {K_{DR}}}} & 0 & 0 & 0 \\
\frac{1}{\varepsilon_5}f_{5V} & 0 & 0 & 0 & \frac{1}{\varepsilon_5}f_{5h} & 0 & 0\\
\frac{1}{\varepsilon_6}f_{6V} & 0 & \frac{1}{\varepsilon_6}f_{6s} & 0 & 0 & \frac{1}{\varepsilon_6}f_{6\mathrm{Ca_i}} & 0 \\
 g_{1V} & 0 & 0 & 0 & 0 & 0 &  g_{1n}
\end{pmatrix}.
\end{equation}
The determinant of $J_{SL}$ is expressed as:
\begin{equation}
    \begin{array}{rcl}
  \varepsilon_1 \varepsilon_2 \varepsilon_3 \varepsilon_4 \varepsilon_5 \varepsilon_6  \det J_{SL} &=& 
     f_{1V} f_{2m_{\mathrm {NaP}}} f_{3s} f_{4m_{\mathrm {K_{DR}}}} f_{5h} f_{6\mathrm{Ca_i}} g_{1n} \vspace{0.05in}\\
    &&- f_{1m_{\mathrm {NaP}}} f_{2V} f_{3s} f_{4m_{\mathrm {K_{DR}}}} f_{5h} f_{6\mathrm{Ca_i}} g_{1n} \vspace{0.05in}\\
   &&-f_{1s}f_{2m_{\mathrm {NaP}}} f_{3V} f_{4m_{\mathrm {K_{DR}}}} f_{5h} f_{6\mathrm{Ca_i}} g_{1n} \vspace{0.05in}\\
   &&- f_{1m_{\mathrm {K_{DR}}}} f_{2m_{\mathrm {NaP}}} f_{3s} f_{4V} f_{5h} f_{6\mathrm{Ca_i}} g_{1n}  \vspace{0.05in}\\
    &&-f_{1h} f_{2m_{\mathrm {NaP}}} f_{3s} f_{4m_{\mathrm {K_{DR}}}} f_{5V} f_{6\mathrm{Ca_i}} g_{1n}  \vspace{0.05in}\\
    &&- f_{1n} f_{2m_{\mathrm {NaP}}} f_{3s} f_{4m_{\mathrm {K_{DR}}}} f_{5h} f_{6\mathrm{Ca_i}} g_{1V}.
    \end{array}
\end{equation}
Our numerical computations show that the set \eqref{eq:fold-Mss-theta} is empty, implying \eqref{eq:slow-layer-theta} has no saddle-node bifurcations. 

Another generic bifurcation of interest is the Hopf bifurcation of the slow layer problem \eqref{eq:slow-layer-theta}, which also leads to a breakdown of GSPT. Due to the high dimensionality of the Jacobian matrix $J_{SL}$, it is challenging to derive an explicit condition for Hopf bifurcations. Numerical continuation reveals that system \eqref{eq:slow-layer-theta} possesses four Hopf bifurcations, but only one occurs within the physiological domain of $q$ and is $O(\varepsilon)$-close to the CDH singularity, the intersection of $M_{ss}$ and $L_s$. As shown in Fig. \ref{fig:twopar-hopf-theta}B, this Hopf bifurcation (red curve) converges to the CDH singularity (yellow curve) as $r_\varepsilon \rightarrow \infty$ (i.e., $\varepsilon \rightarrow 0$). Referred to as a delayed Hopf bifurcation (DHB), it divides $M_{ss}$ into attracting and repelling branches, $M_{ss}^a$ and $M_{ss}^r$. Along $M_{ss}$, the superslow dynamics is obtained by taking the singular limit $\delta \rightarrow 0$ in the superslow system \eqref{eq:superslow-theta}, yielding the 1D \textit{superslow reduced problem}
\begin{equation} \label{eq:ssl-reduced-theta}
\begin{array}{rcl}
     \frac{dq}{dt_{ss}}&=& g_2(\mathrm{Ca_i}, q).
\end{array}    
\end{equation}
According to GSPT, near the normal hyperbolic portions of $M_{ss}$, the flow of \eqref{eq:slow-sys-appendix} is $O(\delta)$-close to that of the superslow reduced problem \eqref{eq:ssl-reduced-theta}.

\subsection{$\varepsilon \rightarrow 0, \, \delta \rightarrow 0$ double singular limit and the CDH singularity} \label{sec:double-singular-theta}

\begin{figure}[!htp]
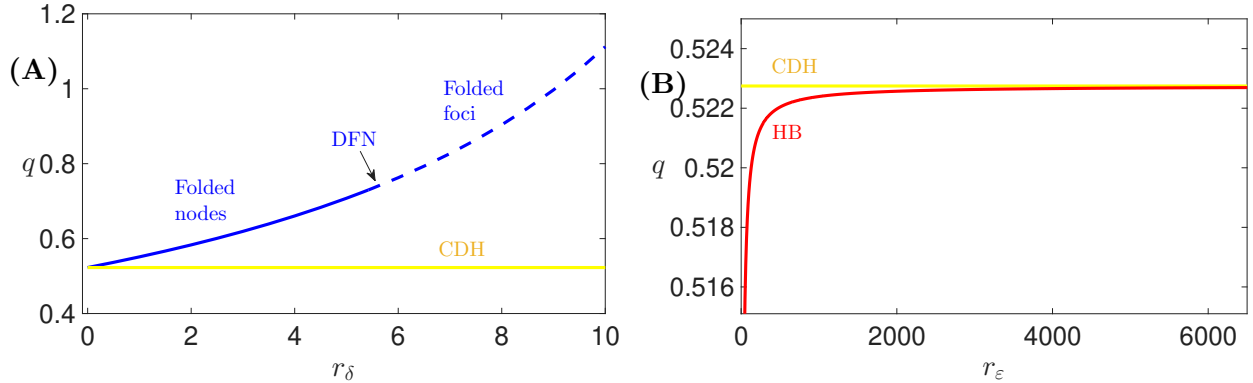

\begin{center}
\begin{tabular}
{@{}p{0.48\linewidth}@{\quad}p{0.48\linewidth}@{}}
\subfigimg[width=\linewidth]{\bfseries{\small{(A)}}}{twopar_folded_singularity.eps} &
\subfigimg[width=\linewidth]{\bfseries{\small{(B)}}}{twopar_hopf.eps} 
\end{tabular}
\end{center}
\caption{The CDH represents the interaction between the canard and DHB mechanisms. (A) The folded singularities (blue curve) of \eqref{eq:desingularization-theta} converge to the CDH (yellow curve) as $r_\delta \rightarrow 0$ (equivalently, $\delta \rightarrow 0$). The blue curve consists of folded nodes (solid) and folded foci (dashed), separated by a degenerate folded node (DFN). (B) The DHB (red curve) of \eqref{eq:slow-layer-theta} converges to the CDH (yellow curve) as $r_\varepsilon \rightarrow \infty$ (equivalently, $\varepsilon \rightarrow 0$).
} 
\label{fig:twopar-hopf-theta}
\end{figure}

We observe that the slow reduced problem \eqref{eq:slow-reduced-theta} and the slow layer problem \eqref{eq:slow-layer-theta} are still singularly perturbed problems with small perturbation parameters. By either taking $\delta \rightarrow 0$ in \eqref{eq:slow-reduced-theta} or taking $\varepsilon \rightarrow 0$ in \eqref{eq:slow-layer-theta}, we obtain the same \textit{slow reduced layer problem}
\begin{equation}\label{eq:slow-reduced-layer-theta}
\begin{array}{rcl}
    \frac{dn}{dt_s}&=&  g_1(V,n),
\end{array} 
\end{equation}
which describes the slow dynamics on $M_s$ while the superslow variable $q$ is fixed. The system \eqref{eq:slow-reduced-layer-theta} interprets the interaction between the 6-fast/2-slow (in Section \ref{sub:eps-viewpoint-theta}) and 7-slow/1-superslow analyses (in Section \ref{sub:delta-viewpoint-theta}). To complete our analysis, we examine the desingularized system associated with the slow reduced layer \eqref{eq:slow-reduced-layer-theta}. This system  can be obtained by taking $\delta \rightarrow 0$ in \eqref{eq:desingularization-theta} to yield
\begin{equation}\label{eq:desingularized-double-limit-theta}
    \begin{array}{rcl}
        \frac{dV}{dt_{d}} &=&  F_0(V, m_{\mathrm {NaP}}, s, m_{\mathrm {K_{DR}}}, h, \mathrm{Ca_i}, n, q, 0) \vspace{0.05in} \\
        \frac{dn}{dt_{d}} &=&\det(\mathbf{J})  g_1.
    \end{array}
\end{equation}
The ordinary singularities of \eqref{eq:desingularized-double-limit-theta} are relaxed to be the superslow manifold $M_{ss}$, while the folded singularities are $\delta \rightarrow 0$ limit of $M_\delta$ in \eqref{eq:folded-singu-orig-theta}:
\begin{equation} 
\begin{array}{rcl}
    \mathcal{M}_0&:=&\{ (V, m_{\mathrm {NaP}}, s, m_{\mathrm {K_{DR}}}, h, \mathrm{Ca_i}, n, q) \in L_s: F_0(V, m_{\mathrm {NaP}}, s, m_{\mathrm {K_{DR}}}, h, \mathrm{Ca_i}, n, q, 0) =0\} \vspace{0.05in}\\
    &=& \{ (V, m_{\mathrm {NaP}}, s, m_{\mathrm {K_{DR}}}, h, \mathrm{Ca_i}, n, q) \in L_s:g_1(V, n) = 0 \}.
    \end{array}
\end{equation}
Thus, in the double singular limit, $\mathcal{M}_0$ becomes a CDH singularity, consistent with the convergence of $\mathcal{M}_\delta$ to the CDH as $\delta \rightarrow 0$, as demonstrated in Fig.~\ref{fig:twopar-hopf-theta}A. The CDH therefore serves as an organizing center for both geometric viewpoints: it is the $\varepsilon \rightarrow 0$ limit of the DHB and the $\delta \rightarrow 0$ limit of the folded singularity $\mathcal{M}_\delta$ (see Fig. \ref{fig:twopar-hopf-theta}). Geometrically, this singularity has been shown to play a crucial role in organizing MMOs within the three-timescale setting \cite{Letson2017, Vo2013, Phan2023}. However, the mere presence of the CDH does not guarantee the occurrence of MMOs \cite{Phan2023}.

\end{appendix}

\bibliographystyle{plain} 
\bibliography{refs.bib} 

\begin{thebibliography}{10}

\bibitem{Avitabile2022}
D.~Avitabile, M.~Desroches, and B.~G. Ermentrout.
\newblock Cross-scale excitability in networks of quadratic integrate-and-fire neurons.
\newblock {\em PLoS Comput. Biol.}, 18(10):e1010569, 2022.

\bibitem{Awal2023}
N.~M. Awal, I.~R. Epstein, T.~J. Kaper, and T.~Vo.
\newblock Symmetry-breaking rhythms in coupled, identical fast–slow oscillators.
\newblock {\em Chaos}, 33(1), 2023.

\bibitem{Baer1989}
S.~M. Baer, T.~Erneux, and J.~Rinzel.
\newblock The slow passage through a {H}opf bifurcation: {D}elay, memory effects, and resonance.
\newblock {\em SIAM J. Appl. Math.}, 49(1):55--71, 1989.

\bibitem{barrio2024dynamics}
Roberto Barrio, Santiago Ib{\'a}{\~n}ez, Jorge~A Jover-Galtier, {\'A}lvaro Lozano, M~{\'A}ngeles Mart{\'\i}nez, Ana Mayora-Cebollero, Carmen Mayora-Cebollero, Luc{\'\i}a P{\'e}rez, Sergio Serrano, and Rub{\'e}n Vigara.
\newblock Dynamics of excitable cells: spike-adding phenomena in action.
\newblock {\em SeMA Journal}, 81(1):113--146, 2024.

\bibitem{barrio2020spike}
Roberto Barrio, Santiago Ib{\'a}{\~n}ez, Luc{\'\i}a P{\'e}rez, and Sergio Serrano.
\newblock Spike-adding structure in fold/hom bursters.
\newblock {\em Communications in Nonlinear Science and Numerical Simulation}, 83:105100, 2020.

\bibitem{Bat2021}
S.~Battaglin and M.~G. Pedersen.
\newblock Geometric analysis of mixed-mode oscillations in a model of electrical activity in human beta-cells.
\newblock {\em Nonlinear Dyn.}, 104(4):4445--4457, 2021.

\bibitem{Beims2018}
M.~W. Beims and J.~A.~C. Gallas.
\newblock Predictability of the onset of spiking and bursting in complex chemical reactions.
\newblock {\em Phys. Chem. Chem. Phys.}, 20(27):18539--18546, 2018.

\bibitem{Bertram1995}
R.~Bertram, M.~J. Butte, T.~Kiemel, and A.~Sherman.
\newblock Topological and phenomenological classification of bursting oscillations.
\newblock {\em Bulletin of mathematical biology}, 57(3):413--439, 1995.

\bibitem{Bertram2008}
R.~Bertram, J.~Rhoads, and W.~P. Cimbora.
\newblock A phantom bursting mechanism for episodic bursting.
\newblock {\em Bulletin of Mathematical Biology}, 70:1979--1993, 2008.

\bibitem{carracedo2013neocortical}
Lucy~M Carracedo, Henrik Kjeldsen, Leonie Cunnington, Alastair Jenkins, Ian Schofield, Mark~O Cunningham, Ceri~H Davies, Roger~D Traub, and Miles~A Whittington.
\newblock A neocortical delta rhythm facilitates reciprocal interlaminar interactions via nested theta rhythms.
\newblock {\em Journal of Neuroscience}, 33(26):10750--10761, 2013.

\bibitem{Curtu2010}
R.~Curtu.
\newblock Singular {H}opf bifurcation and mixed-mode oscillations in a two-cell inhibitory neural network.
\newblock {\em Phys. D: Nonlinear Phenom.}, 239(9):504--514, 2010.

\bibitem{CR2011}
R.~Curtu and J.~Rubin.
\newblock Interaction of canard and singular {H}opf mechanisms in a neural model.
\newblock {\em SIAM J. Appl. Dyn. Syst.}, 10(4):1443--1479, 2011.

\bibitem{Maess2014}
P.~{{D}e Maesschalck}, E.~Kutafina, and N.~Popovi{\'c}.
\newblock Three time-scales in an extended {B}onhoeffer–van der {P}ol oscillator.
\newblock {\em J. Dyn. Differ. Equ.}, 26:955--987, 2014.

\bibitem{Desroches2012}
M.~Desroches, J.~Guckenheimer, B.~Krauskopf, C.~Kuehn, H.~M. Osinga, and M.~Wechselberger.
\newblock Mixed-mode oscillations with multiple time scales.
\newblock {\em SIAM Rev.}, 54(2):211--288, 2012.

\bibitem{Desroches2013}
M.~Desroches, T.~J. Kaper, and M.~Krupa.
\newblock Mixed-mode bursting oscillations: Dynamics created by a slow passage through spike-adding canard explosion in a square-wave burster.
\newblock {\em Chaos}, 23(4), 2013.

\bibitem{DK2018}
M.~Desroches and V.~Kirk.
\newblock Spike-adding in a canonical three-time-scale model: superslow explosion and folded-saddle canards.
\newblock {\em SIAM J. Appl. Dyn. Syst.}, 17(3):1989--2017, 2018.

\bibitem{Desroches2022}
M.~Desroches, J.~Rinzel, and S.~Rodrigues.
\newblock Classification of bursting patterns: A tale of two ducks.
\newblock {\em PLoS Comput. Biol.}, 18(2), 2022.

\bibitem{doedel1981auto}
Eusebius~J Doedel.
\newblock Auto: A program for the automatic bifurcation analysis of autonomous systems.
\newblock {\em Congr. Numer}, 30(265-284):25--93, 1981.

\bibitem{doedel1997auto97}
Eusebius~J Doedel.
\newblock Auto97: Continuation and bifurcation software for ordinary differential equations (with homcont).
\newblock {\em Technical Report, Concordia University}, 1997.

\bibitem{engler2026delays}
Hans Engler, Hans Kaper, Tasso Kaper, and Theodore Vo.
\newblock Delays and advances in the onset of instability in the shishkova equation.
\newblock {\em Quarterly of Applied Mathematics}, 84(2):325--349, 2026.

\bibitem{Ermentrout2010}
B.~Ermentrout and D.~H. Terman.
\newblock {\em Mathematical foundations of neuroscience}, volume~35.
\newblock Springer, 2010.

\bibitem{Farjami2020}
S.~Farjami, R.~P.~D. Alexander, D.~Bowie, and A.~Khadra.
\newblock Bursting in cerebellar stellate cells induced by pharmacological agents: Non-sequential spike adding.
\newblock {\em PLoS Comput. Biol.}, 16(22):e1008463, 2020.

\bibitem{Fenichel1979}
N.~Fenichel.
\newblock Geometric singular perturbation theory for ordinary differential equations.
\newblock {\em J. Differ. Equ.}, 31(1):53--98, 1979.

\bibitem{fernandez1997isolas}
F~Fern{\'a}ndez-S{\'a}nchez, E~Freire, and AJ~Rodr{\'\i}guez-Luis.
\newblock Isolas, cusps and global bifurcations in an electronic oscillator.
\newblock {\em Dynamics and Stability of Systems}, 12(4):319--336, 1997.

\bibitem{Golubitsky2001}
M.~Golubitsky, K.~Josic, and T.~J. Kaper.
\newblock An unfolding theory approach to bursting in fast-slow systems.
\newblock {\em Global analysis of dynamical systems}, pages 277--308, 2001.

\bibitem{Harvey2011}
E.~Harvey, V.~Kirk, M.~Wechselberger, and J.~Sneyd.
\newblock Multiple timescales, mixed mode oscillations and canards in models of intracellular calcium dynamics.
\newblock {\em J. Nonlinear Sci.}, 21:639--683, 2011.

\bibitem{Hayes2016}
M.~G. Hayes, T.~J. Kaper, P.~Szmolyan, and M.~Wechselberger.
\newblock Geometric desingularization of degenerate singularities in the presence of fast rotation: {A} new proof of known results for slow passage through {H}opf bifurcations.
\newblock {\em Indag. Math.}, 27(5):1184--1203, 2016.

\bibitem{he2026complex}
Ke~He, Sue~Ann Campbell, and Shenquan Liu.
\newblock Complex dynamics induced by multiple timescales in a wilson--cowan model with homeostatic plasticity.
\newblock {\em Chaos: An Interdisciplinary Journal of Nonlinear Science}, 36(1), 2026.

\bibitem{Hindmarsh1984}
J.~L. Hindmarsh and R.~M. Rose.
\newblock A model of neuronal bursting using three coupled first order differential equations.
\newblock {\em Proceedings of the Royal society of London. Series B. Biological sciences}, 221(1222):87--102, 1984.

\bibitem{Hudson1979}
J.~L. Hudson, M.~Hart, and D.~Marinko.
\newblock An experimental study of multiple peak periodic and nonperiodic oscillations in the {B}elousov--{Z}habotinskii reaction.
\newblock {\em J. Chem. Phys.}, 71(4):1601--1606, 1979.

\bibitem{Izhikevich2000}
E.~M. Izhikevich.
\newblock Neural excitability, spiking and bursting.
\newblock {\em International journal of bifurcation and chaos}, 10(06):1171--1266, 2000.

\bibitem{Kepecs2000}
A.~Kepecs and X.~J. Wang.
\newblock Analysis of complex bursting in cortical pyramidal neuron models.
\newblock {\em Neurocomputing}, 32:181--187, 2000.

\bibitem{Kimrey2020}
J.~Kimrey, T.~Vo, and R.~Bertram.
\newblock Big ducks in the heart: canard analysis can explain large early afterdepolarizations in cardiomyocytes.
\newblock {\em SIAM J. Appl. Dyn. Syst.}, 19(3):1701--1735, 2020.

\bibitem{2ndKimrey2020}
J.~Kimrey, T.~Vo, and R.~Bertram.
\newblock Canard analysis reveals why a large $\mathrm {C}a^{2+}$ window current promotes early afterdepolarizations in cardiac myocytes.
\newblock {\em PLoS Comput. Biol.}, 16(11):e1008341, 2020.

\bibitem{Kingni2015}
S.~T. Kingni, B.~Nana, G.~M. Ngueuteu, P.~Woafo, and J.~Danckaert.
\newblock Bursting oscillations in a 3d system with asymmetrically distributed equilibria: mechanism, electronic implementation and fractional derivation effect.
\newblock {\em Chaos, Solitons $\&$ Fractals}, 71:29--40, 2015.

\bibitem{koksal2020canard}
Elif K{\"o}ksal~Ers{\"o}z, Mathieu Desroches, Antoni Guillamon, John Rinzel, and Joel Tabak.
\newblock Canard-induced complex oscillations in an excitatory network.
\newblock {\em Journal of mathematical biology}, 80(7):2075--2107, 2020.

\bibitem{Krupa2008a}
M.~Krupa, N.~Popovi{\'c}, N.~Kopell, and H.~G. Rotstein.
\newblock Mixed-mode oscillations in a three time-scale model for the dopaminergic neuron.
\newblock {\em Chaos}, 18(1):015106, 2008.

\bibitem{Krupa2012}
M.~Krupa, A.~Vidal, M.~Desroches, and F.~Cl{\'e}ment.
\newblock Mixed-mode oscillations in a multiple time scale phantom bursting system.
\newblock {\em SIAM J. Appl. Dyn. Syst.}, 11(4):1458--1498, 2012.

\bibitem{Kugler2016}
P.~K{\"u}gler.
\newblock Early afterdepolarizations with growing amplitudes via delayed subcritical hopf bifurcations and unstable manifolds of saddle foci in cardiac action potential dynamics.
\newblock {\em PLoS One}, 11(3):p.e0151178, 2016.

\bibitem{Kugler2018}
P.~K{\"u}gler, A.~H. Erhardt, and M.~A.~K. Bulelzai.
\newblock Early afterdepolarizations in cardiac action potentials as mixed mode oscillations due to a folded node singularity.
\newblock {\em PLoS One}, 13(12):e0209498, 2018.

\bibitem{Letson2017}
B.~Letson, J.~E. Rubin, and T.~Vo.
\newblock Analysis of interacting local oscillation mechanisms in three-timescale systems.
\newblock {\em SIAM J. Appl. Dyn. Syst.}, 77(3):1020--1046, 2017.

\bibitem{Leutcho2020b}
G.~D. Leutcho, J.~Kengne, L.~K. Kengne, A.~Akgul, V.~T. Pham, and S.~Jafari.
\newblock A novel chaotic hyperjerk circuit with bubbles of bifurcation: mixed-mode bursting oscillations, multistability, and circuit realization.
\newblock {\em Physica Scripta}, 95(7):075216, 2020.

\bibitem{Leutcho2020a}
G.~D. Leutcho, J.~Kengne, A.~Ngoumkam Negou, T.~Fonzin Fozin, V.~T. Pham, and S.~Jafari.
\newblock A modified simple chaotic hyperjerk circuit: coexisting bubbles of bifurcation and mixed-mode bursting oscillations.
\newblock {\em Zeitschrift f{\"u}r Naturforschung A}, 75(7):593--607, 2020.

\bibitem{Ma2022}
X.~Ma, Q.~Bi, and L.~Wang.
\newblock Complex periodic bursting structures in the rayleigh--van der pol--duffing oscillator.
\newblock {\em J. Nonlinear Sci.}, 32(2):25, 2022.

\bibitem{Nan2015}
P.~Nan, Y.~Wang, V.~Kirk, and J.~E. Rubin.
\newblock Understanding and distinguishing three-time-scale oscillations: {C}ase study in a coupled {M}orris-{L}ecar system.
\newblock {\em SIAM J. Appl. Dyn. Syst.}, 14(3):1518--1557, 2015.

\bibitem{Neishtadt1987}
A.~Neishtadt.
\newblock On delayed stability loss under dynamical bifurcations {I}.
\newblock {\em Differ. Equ.}, 23:1385--1390, 1987.

\bibitem{Neishtadt1988}
A.~Neishtadt.
\newblock On delayed stability loss under dynamical bifurcations {II}.
\newblock {\em Differ. Equ.}, 24:171--176, 1988.

\bibitem{Nowacki2012}
J.~Nowacki, H.~M. Osinga, and K.~Tsaneva-Atanasova.
\newblock Dynamical systems analysis of spike-adding mechanisms in transient bursts.
\newblock {\em J. Math. Neurosci.}, 2:1--28, 2012.

\bibitem{Organ2003}
L.~Organ, I.~Z. Kiss, and J.~L. Hudson.
\newblock Bursting oscillations during metal electrodissolution: Experiments and model.
\newblock {\em J. Phys. Chem. B}, 107(27):6648--6659, 2003.

\bibitem{Pavlidis2022}
E.~Pavlidis, F.~Campillo, A.~Goldbeter, and M.~Desroches.
\newblock Multiple-timescale dynamics, mixed mode oscillations and mixed affective states in a model of bipolar disorder.
\newblock {\em Cognitive Neurodynamics}, pages 1--19, 2022.

\bibitem{Phan2023}
N.~A. Phan and Y.~Wang.
\newblock Mixed-mode oscillations in a three-timescale coupled morris-lecar system.
\newblock {\em Chaos}, 34(5):053119, 2024.

\bibitem{Pittman2021}
B.~R. Pittman-Polletta, Y.~Wang, D.~A. Stanley, C.~E. Schroeder, M.~A. Whittington, and N.~J. Kopell.
\newblock Differential contributions of synaptic and intrinsic inhibitory currents to speech segmentation via flexible phase-locking in neural oscillators.
\newblock {\em PLoS Comput. Biol.}, 17(4):p.e1008783, 2021.

\bibitem{Rinzel1987}
J.~Rinzel.
\newblock A formal classification of bursting mechanisms in excitable systems.
\newblock {\em Mathematical Topics in Population Biology, Morphogenesis and Neurosciences: Proceedings of an International Symposium held in Kyoto, November 10--15, 1985}, pages 267--281, 1987.

\bibitem{Rinzel2006}
J.~Rinzel.
\newblock Bursting oscillations in an excitable membrane model.
\newblock {\em Ordinary and Partial Differential Equations: Proceedings of the Eighth Conference held at Dundee, Scotland, June 25–29, 1984}, pages 304--316, 2006.

\bibitem{RE1998}
J.~Rinzel and G.~B. Ermentrout.
\newblock Analysis of neural excitability and oscillations.
\newblock {\em Methods in neuronal modeling}, 2:251--292, 1998.

\bibitem{Rinzel1982}
J.~Rinzel and W.~C. Troy.
\newblock Bursting phenomena in a simplified oregonator flow system model.
\newblock {\em J. Chem. Phys.}, 76(4):1775--1789, 1982.

\bibitem{Rubin2017}
J.~E. Rubin, J.~Signerska-Rynkowska, J.~D. Touboul, and A.~Vidal.
\newblock Wild oscillations in a nonlinear neuron model with resets: (ii) mixed-mode oscillations.
\newblock {\em Discrete Contin. Dyn. Syst.}, 22(10):4003--4039, 2017.

\bibitem{sanders2007averaging}
Jan~A Sanders, Ferdinand Verhulst, and James Murdock.
\newblock {\em Averaging methods in nonlinear dynamical systems}, volume~59.
\newblock Springer, 2007.

\bibitem{SW2001}
P.~Szmolyan and M.~Wechselberger.
\newblock Canards in $\mathbb{R}^3$.
\newblock {\em J. Differ. Equ.}, 177(2):419--453, 2001.

\bibitem{Teka2012}
W.~Teka, J.~Tabak, and R.~Bertram.
\newblock The relationship between two fast/slow analysis techniques for bursting oscillations.
\newblock {\em Chaos}, 22(4):043117, 2012.

\bibitem{tsaneva2010full}
Krasimira Tsaneva-Atanasova, Hinke~M Osinga, Thorsten Rie{\ss}, and Arthur Sherman.
\newblock Full system bifurcation analysis of endocrine bursting models.
\newblock {\em Journal of theoretical biology}, 264(4):1133--1146, 2010.

\bibitem{Vo2010}
T.~Vo, R.~Bertram, J.~Tabak, and M.~Wechselberger.
\newblock Mixed mode oscillations as a mechanism for pseudo-plateau bursting.
\newblock {\em J. Comput. Neurosci.}, 28:443--458, 2010.

\bibitem{Vo2013}
T.~Vo, R.~Bertram, and M.~Wechselberger.
\newblock Multiple geometric viewpoints of mixed mode dynamics associated with pseudo-plateau bursting.
\newblock {\em SIAM J. Appl. Dyn. Syst.}, 12(2):789--830, 2013.

\bibitem{Vo2014}
T.~Vo, J.~Tabak, R.~Bertram, and M.~Wechselberger.
\newblock A geometric understanding of how fast activating potassium channels promote bursting in pituitary cells.
\newblock {\em J. Comput. Neurosci.}, 36:259--278, 2014.

\bibitem{vo2012bifurcations}
Theodore Vo, Richard Bertram, and Martin Wechselberger.
\newblock Bifurcations of canard-induced mixed mode oscillations in a pituitary lactotroph model.
\newblock {\em Discrete Contin Dyn Syst}, 32(8):2879--2912, 2012.

\bibitem{vo2026symmetric}
Theodore Vo and Yangyang Wang.
\newblock A symmetric mechanism for symmetry-breaking in oscillator networks with strong nonlinear coupling.
\newblock {\em arXiv preprint arXiv:2606.17780}, 2026.

\bibitem{WR2016}
Y.~Wang and J.~E. Rubin.
\newblock Multiple timescale mixed bursting dynamics in a respiratory neuron model.
\newblock {\em J. Comput. Neurosci.}, 41:245--268, 2016.

\bibitem{WR2017}
Y.~Wang and J.~E. Rubin.
\newblock Timescales and mechanisms of sigh-like bursting and spiking in models of rhythmic respiratory neurons.
\newblock {\em J. Math. Neurosci.}, 7:1--39, 2017.

\bibitem{WR2020}
Y.~Wang and J.~E. Rubin.
\newblock Complex bursting dynamics in an embryonic respiratory neuron model.
\newblock {\em Chaos}, 30(4):043127, 2020.

\bibitem{wang2026dynamical}
Yangyang Wang and Benjamin~R Pittman-Polletta.
\newblock Dynamical mechanisms of flexible phase-locking in cortical theta oscillators.
\newblock {\em arXiv preprint arXiv:2605.08014}, 2026.

\bibitem{Wechselberger2005}
M.~Wechselberger.
\newblock Existence and bifurcation of canards in $\mathbb{R}^3$ in the case of a folded node.
\newblock {\em SIAM J. Appl. Dyn. Syst.}, 4(1):101--139, 2005.

\bibitem{Wechselberger2012}
M.~Wechselberger.
\newblock A propos de canards (apropos canards).
\newblock {\em Transactions of the American Mathematical Society}, 364(6):3289--3309, 2012.

\bibitem{Yu2008}
N.~Yu, R.~Kuske, and Y.~X. Li.
\newblock Stochastic phase dynamics and noise-induced mixed-mode oscillations in coupled oscillators.
\newblock {\em Chaos}, 18(1), 2008.

\bibitem{Yu2021}
N.~Yu, X.~Xia, and J.~Liyau.
\newblock Mixed mode bursting oscillations induced by birhythmicity and noise.
\newblock {\em arXiv preprint arXiv:2110.07167}, 2021.

\end{thebibliography}
\end{document}